\documentclass[a4paper,11pt]{article}
\pdfoutput=1 
\usepackage{jheppub} 

  \usepackage{etoolbox}
  \patchcmd{\maketitle}{\@fpheader}{}{}{}

\usepackage[T1]{fontenc} 
\usepackage[utf8]{inputenc} 
\usepackage{microtype} 
\usepackage{mdframed} 
\usepackage{graphicx}
\usepackage{todonotes} 

\usepackage{amsmath}
\usepackage{amsfonts}
\usepackage{amssymb}
\usepackage{stmaryrd}
\usepackage{mathtools}
\usepackage{mathrsfs}
\usepackage{mleftright} \mleftright
\usepackage{tensor} 
\usepackage{braket} 
\usepackage{mathrsfs} 
\usepackage{bbold}
\usepackage{color}
\usepackage{braket}
\usepackage{slashed}
\usepackage{hyperref}

\newcommand{\comment}[1]{}

\DeclareMathAlphabet{\mathfs}{U}{rsfs}{m}{n}                     %

\newcommand{\inter}{{\lrcorner}}
\newcommand{\mfs}[1]{\mathfs {#1}}                               %
\newcommand{\sL}{{\mfs L}}
\newcommand{\Lie}{\sL}

\newcommand{\n}{\nonumber}
\newcommand{\be}{\nopagebreak[3]\begin{equation}}
\newcommand{\ee}{\end{equation}}
\newcommand{\bee}{\nopagebreak[3]\begin{equation*}}
\newcommand{\eee}{\end{equation*}}
\newcommand{\ba}{\nopagebreak[3]\begin{eqnarray}}
\newcommand{\ea}{\end{eqnarray}}
\newcommand{\baa}{\nopagebreak[3]\begin{eqnarray*}}
\newcommand{\eaa}{\end{eqnarray*}}

\title{Surface Charges for Gravity and Electromagnetism\\ in the First Order Formalism}

\author[1]{Ernesto Frodden}
\author[1,2]{and Diego Hidalgo}

\affiliation[1]{Centro de Estudios Cient\'ificos (CECs), Av. Arturo Prat 514, Valdivia, Chile}
\affiliation[2]{Departamento de F\'isica, Universidad de Concepci\'on, Casilla 160-C, Concepci\'on, Chile}

\emailAdd{efrodden@cecs.cl}
\emailAdd{dhidalgo@cecs.cl}

\preprint{CECS-PHY-17/05}

\abstract{
A new derivation of surface charges for 3+1 gravity coupled to Electromagnetism is obtained. Gravity theory is written in the tetrad-connection variables.
The general derivation starts from the Lagrangian and uses the covariant symplectic formalism in the language of forms. For gauge theories surface charges disentangle physical from gauge symmetries through the use of Noether identities and the exactness symmetry condition. The surface charges are quasilocal, explicitly coordinate independent, gauge invariant, and background independent. For a black hole family solution the surface charge conservation implies the first law of black hole mechanics. As a check we show the first law for black hole electrically charged, rotating, and with an asymptotically constant curvature (the Kerr-Newman (anti-)de Sitter family). The charges, including the would-be mass term appearing in the first law, are quasilocal. It is not required a reference to the asymptotic structure of the spacetime nor boundary conditions, and therefore topological terms do not play a r\^ole. Finally, surface charges formulae for Lovelock gravity coupled to Electromagnetism are exhibited. It generalizes the one derived in a recent work by G. Barnich, P. Mao, and R. Ruzziconi. The two different symplectic methods to define surface charges are compared and shown equivalent.  
}

\begin{document}

\maketitle
\flushbottom

\section{Introduction}
\label{sec:introduction}

To find the quantum degrees of freedom responsible for the black hole entropy remains one of the main questions that fuels the research of a quantum theory of gravity. The semiclassical analysis, the study of quantum field theory on fixed black hole spacetimes, ensures that the entropy is proportional to one-fourth of the horizon area (in units where $G=c=1$). 

Because the entropy corresponds to the area of the horizon, the expectation is that the microscopic degrees of freedom responsible for the entropy are localized around the horizon itself. At least two prevailing approaches construct quantum horizons models dwelling on this later view. 

A first one uses symplectic methods to build a Chern-Simons description of the horizon \cite{Ashtekar:2000eq,Engle:2009vc,Engle:2010kt,Perez:2010pq} in the context of loop quantum gravity. A second one uses tools \cite{Brown:1986nw,Barnich:2001jy} developed in the context of holography, specifically asymptotic symmetries and associated charges, to describe the black hole horizon (some examples are \cite{Strominger:1997eq,Guica:2008mu,Compere:2015mza,Afshar:2016uax,Hawking:2016msc}). Both approaches start from a classical description of the black hole horizon through boundary conditions that left some freedom for the variables at the boundary. In both cases one may wonder if in that freedom there are true degrees of freedom. Surprisingly, in certain context they have been called would-be gauge degrees of freedom \cite{Carlip:1995zj,Geiller:2017xad} as some of them would be degrees of freedom that come from the partial freezing of the gauge symmetries at the boundary. That seems artificial as they are strongly dependent on the particular choice of boundary conditions.
	
With this context in mind, the present work is a first step to explore, from a different perspective, the common basis where both approaches above are standing on. That is, the study of physical and gauge symmetries of the symplectic structure when boundary conditions are imposed. To do so, we focus on {\it surface charges}, motivated in subsection \ref{motsurfacecharges}, as  a main quantity that encodes the physical information related with symmetries, in the context of gauge theories. 

As a result we rederive explicitly the general expression of surface charges from the covariant symplectic method in the language of forms, section \ref{surfacecharges}. Then, to describe gravity, we choose to depart from the usual metric approach by using the tetradic-connection formalism. We also deal directly with the more general case of a gravity theory (with cosmological constant) coupled to Electromagnetism in four dimensions. That allows us to obtain, as a second and main result, compact formulae for the surface charges that are gauge invariant and do not require reference to a coordinate system, equation (\ref{kGREM}).  A valuable conclusion of this work is an explicit exhibition of surface charges formula and the tight requirement for them to satisfy a conservation law. This paves the way to disentangle physical from gauge symmetries in the context of asymptotic symmetries and boundary conditions previuosly discussed.

The content of these notes goes as follow: In section \ref{surfacecharges} we rederive surface charges for a general gauge theory. In sections \ref{generalrelativity}-\ref{generalelectro} we progressively establish the explicit formulae of the surface charges for the theory of gravity coupled to Electromagnetism. In subsection \ref{topo} we show that boundary terms in the Lagrangian have not effects on surface charges.
In subsection \ref{bhexample} we perform a test of the reliability of the formalism by recovering the standard first law of black hole mechanics in a quasilocal way. In section \ref{dworld} we study the generalization to an arbitrary dimension. We compute the surface charges formula for Lovelock gravity coupled to Electromagnetism. Finally, in the appendix \ref{appendix} we further present a general comparison of the canonical covariant symplectic approach with other covariant techniques inherited from the BRST formalism \cite{Barnich:2016rwk}.

\subsection {Why surface charges?}
\label{motsurfacecharges}

It is not enough well-known that the first Noether theorem does not apply for gauge symmetries. The reason is that the would-be Noether current is trivially conserved, i.e., it is conserved without requiring the equations of motion (for more discussion see \cite{Compere:2007az}). As we will show, this is in fact a consequence of the second Noether theorem. Then, the would-be Noether charge is ambiguous as it has the arbitrary gauge parameter on it. For example, for the electromagnetic field the `gauge current' for the gauge symmetry, $\delta_\lambda A=d\lambda$, is $J_\lambda= d(\lambda * F)$ and the `gauge charge' $Q_\lambda=\oint \lambda*F$ depends explicitly on the arbitrary gauge parameter.

A usual way to cure this lack of meaning for the would-be Noether current is to assume extra structure for the fields and gauge parameters at an asymptotic spacetime region. With this, the charge computed out of the current may acquire a physical meaning. However, still ambiguities related with the action boundary term may affect the value of the current and extra input as differentiability of the action may be needed to have a well defined charge.   

In the case of gravity, the asymptotic structure of flat, de Sitter, or anti-de Sitter spacetimes are drastically different. The boundary term in the action to guarantee differentiability of the action changes for each case. This fact makes the definition of asymptotic charges problematic. A more general approach not resting on the particular asymptotic structure of the spacetime is certainly desirable. 

The quantities known as {\it surface charges} provide the necessary generalization \cite{Barnich:2001jy,Barnich:2007bf}. They are conserved quantities for a physical symmetry in the context of gauge theories.

In the next section we show how to compute surface charges in general, from canonical symplectic methods. The computation relies on the Lagrangian but is not dependent of the ambiguities of the boundary terms that one may add to it.

An interesting property of surface charges is that they are {\it quasilocal}. It is not needed to use an asymptotic spacetime structure to define them. On the other hand, one may perfectly compute them on asymptotic regions too.

It is worth to notice that surface charges are a particular case of a generalization of conserved currents. In \cite{Barnich:2000zw} it is explained that the Noether's first theorem can be rephrased as a cohomology in the context of the BRST symmetry. Then, it is proven that higher order conserved currents are in correspondence with a generalization of `global symmetries'. This is a generalization of the Noether's first theorem. The usual Noether current is the first one of these currents. The surface charges are built out of the second of these currents. To understand the rest of the currents from a canonical symplectic perspective is an interesting problem left for a future work.  

\section{Surface Charges for Gauge Theories}
\label{surfacecharges}

In gauge theories field transformations due to gauge and rigid symmetries are entangled. This can puzzle the definition of physical quantities like charges. On the frame of covariant symplectic methods we can start studying both gauge and rigid symmetries on the same foot, and then to make the difference at a crucial step. We start by considering the Noether procedure for infinitesimal symmetry transformations in the language of forms. We specify this, first to the case of gauge symmetries, and then to the case where diffeomorphism is one of the gauge symmetries. At the end, we will define and assume the existence of exact transformations to produce physically sensible results.  We follow in general lines the canonical covariant symplectic approach \cite{Wald:1999wa,Iyer:1994ys} but having in mind the {\it invariant} symplectic approach used in \cite{Barnich:2001jy,Barnich:2007bf} where surface charges are defined. Another useful reference is section 3 in \cite{Compere:2009dp}, or \cite{Prabhu:2017} where a close general treatment is performed.

Consider a Lagrangian form $L[\Phi]$ for a collection of fields $\Phi$. The arbitrary variation is
\be
\delta L = E(\Phi) \delta \Phi+d\Theta(\delta\Phi),
\ee
with $E(\Phi)=0$ the equations of motion, and $\Theta(\delta\Phi)$ a boundary term. The Lagrangian has a symmetry if for certain infinitesimal variations over the configuration space it becomes at most an exact form
\be
\delta_\epsilon L = dM_\epsilon,
\ee
we call $\epsilon$ the collection of parameters that generate the infinitesimal symmetry, and $\delta_\epsilon$ denotes the infinitesimal transformation generated over any quantity. 
In the case $M_\epsilon\neq 0$, the usual notion of symmetry for the action $S[\Phi]=\int_{\cal M}L[\Phi]$ is recovered by choosing a vanishing of the symmetry parameters at a neighborhood of the boundary of the manifold. This can be done always only for gauge symmetries. 

The fields transform under a symmetry  as $\delta_\epsilon\Phi$, therefore
\ba
dM_\epsilon=E(\Phi)\delta_\epsilon\Phi+d\Theta(\delta_\epsilon\Phi).
\ea
Now, let us assume that the transformation $\delta_\epsilon \Phi$ is linear in the symmetry parameters $\epsilon$. This assumption allows us to make a crucial step. We can remove the derivatives over all symmetry parameters and formally decompose 
\be
E(\Phi)\delta_\epsilon\Phi=dS_\epsilon-N_\epsilon,
\label{noether0}
\ee
such that in $N_\epsilon$ the symmetry parameters appears only as factors. We will use a hat to remember that the equations hold on-shell, for instance $S_\epsilon\ \widehat =\  0$ or $N_\epsilon\ \widehat =\  0$. Using the new expression for $E(\Phi)\delta_\epsilon\Phi$ we obtain
\be
d[\Theta(\delta_\epsilon \Phi)-M_\epsilon+S_\epsilon]=N_\epsilon.
\label{notheridentity}
\ee
Now, we restrict ourselves to gauge symmetries. For them, the very structure of the last equation implies\footnote{$N_\epsilon$ can be factorized by the arbitrary parameters $\epsilon$ and at the same time it is equal to an exact form, this implies that $N_\epsilon$ vanishes. Proof: Integrate (\ref{notheridentity}) and choose the parameters to vanish at a neighborhood of the boundary.} 
\be
N_\epsilon=0,
\ee
these are called Noether identities and there is one of them for each independent gauge parameter. These are the usual constraints of the theory due to the redundancy of using gauge variables and are reason why the First Noether theorem does not apply for gauge symmetries.

Then, it is natural to define the form
\be
J_\epsilon\equiv\Theta(\delta_\epsilon\Phi)-M_\epsilon+S_\epsilon,
\ee
which by virtue of the Noether identities satisfies
\be
dJ_\epsilon=0.
\label{dJ=0}
\ee
Note that the statement is {\it off-shell} and therefore $J_\epsilon$ is not a current. However this quantity reduces on-shell to what is usually called the Noether current $J_\epsilon\ \widehat = \ \Theta(\delta_\epsilon\Phi)-M_\epsilon$. As far as $\delta_\epsilon$ generates a gauge symmetry this current is trivial as its off-shell conservation law suggests. However, with two more ingredients this current generates non-trivial and finite charges. These extra assumptions are that $\delta_\epsilon$ is an {\it exact symmetry} of the fields, i.e., $\delta_\epsilon\Phi=0$ (further discussed in the paragraph before (\ref{kconserved})), and that the boundary term in the Lagrangian is consistent with the boundary conditions \cite{Aros:1999id}. This is the standard Noether procedure which for gauge theories needs to be suplemented with extra information. 

However, there is an alternative. We can follow a quasilocal approach that does not make use of the asymptotic structure. The cost is the relying on a linearized theory. This is shown in the following.

The Poincar\'e lemma ensures that a closed form is locally exact, that is, there exists $\widetilde Q_\epsilon$ such that\footnote{Equations (\ref{dJ=0}) and (\ref{J=dQ}) suggest that non-trivial on-shell currents are those which satisfy a conservation law in the whole spacetime but for which the Poincar\'e lemma can not extended to the whole spacetime, i.e., they correspond to the equivalence classes of closed forms which are not exact, i.e., the de Rham cohomology. We refer to \cite{Barnich:2000zw} for a rephrasing of Noether theorems using the cohomology of the BRST symmetry.} 
\be
J_\epsilon=d\widetilde Q_\epsilon.
\label{J=dQ}
\ee

Now, consider an off-shell variation
\ba
\delta \Theta (\delta_\epsilon \Phi)-\delta M_\epsilon+\delta S_\epsilon & = & d\delta \widetilde Q_\epsilon.
\label{varJ}
\ea
We assume that $\delta d=d\delta$. The double hat will be used to remember that, besides the equations of motion, the linearized equations of motion hold too. For instance $\delta S_\epsilon$ does not vanishes on-shell. We need the extra assumption that $\delta\Phi$ satisfies the linearized equations of motion, $\delta S_\epsilon \ \widehat{\widehat = }\ 0$.

The presymplectic structure density is an antisymmetrized double variation of the fields on the phase space, defined by 
\ba
\Omega(\delta_1,\delta_2)&\widehat{\widehat = }&\delta_1\Theta(\delta_2\Phi)-\delta_2 \Theta(\delta_1\Phi)-\Theta([\delta_1,\delta_2]\Phi),
\label{symplet}
\ea
where the boundary term in the action $\Theta(\delta\Phi)$ is also referred to as the presymplectic potential density. The variations, $\delta\Phi$, are assumed to satisfy the linearized equation of motion. Note that $\Omega(\delta_1,\delta_2)$ is a double variation in the phase space and a $(D-1)$-form in spacetime. The double variation can also be understood as a two-form in the phase space. The last term, $\Theta([\delta_1,\delta_2]\Phi)$, should be considered because variations of fields on the phase space do not commute in general \cite{Green:2013ica}. The prefix in presymplectic stands for the fact that variations $\delta_{1,2}$ on the fields can also be gauge symmetry transformations. We need it because by using gauge variables there is not a systematic way to disentangle the gauge redundancy from the phase space. In this sense the phase space is degenerated. It contains gauge orbits, i.e., family of points identified through gauge transformations. In other words, if $\cal M$ is the manifold where $L[\Phi]$ is defined, then, on-shell, $\int_{\partial\cal M}\Omega(\delta_1,\delta_2)$ has degenerated directions \cite{Wald:1999wa}. Precisely those ones associated to infinitesimal gauge transformations.

Considering the presymplectic structure density evaluated in a gauge variation, $\Omega(\delta,\delta_\epsilon)$, we rewrite (\ref{varJ}) as
\ba
\Omega(\delta,\delta_\epsilon) =   -\delta_\epsilon\Theta(\delta\Phi)-\Theta([\delta,\delta_\epsilon]\Phi)+\delta M_\epsilon-\delta S_\epsilon+ d\delta \widetilde Q_\epsilon,
\ea
To go further let us assume that $\epsilon$ contains diffeomorphisms. More precisely, suppose the collection of gauge parameters can be split as $\epsilon=(\xi,\lambda)$, i.e., $\delta_\epsilon =\delta_\xi+\delta_\lambda$. Such that $\xi$ is a vector field generating infinitesimal diffeomorphisms and $\delta_\lambda$ denotes the rest of infinitesimal gauge symmetry transformations. For a form $\omega$ that is invariant under $\delta_\lambda$, the infinitesimal diffeomorphism transformations are generated through a Lie derivative\footnote{If the form is a gauge variable an ambiguity arises for the Lie derivative and the Cartan formula (\ref{cartanformula}) has to be corrected. We address this point in the examples.} 
\be
\delta_\xi\omega=\Lie_\xi\omega=d(\xi\inter\omega)+\xi\inter d\omega, 
\label{cartanformula}
\ee
we use $\inter$ to denote the interior product over forms. For a vector field $\xi=\xi^\mu\partial_\mu$ and a one-form $\omega=\omega_\mu dx^\mu$, both expressed in coordinate components, the interior product is $\xi\inter\omega=\xi^\mu \omega_\mu$. The interior product distributes over the wedge product of forms exactly as the exterior derivative does. The exterior derivative, $d$, and the interior product, $\inter$, act only on the immediate term at the right of the symbol unless explicit parenthesis are drawn.

We assume the Lagrangian and the presymplectic potential density are left invariant under the transformation generated by $\lambda$,\footnote{Note that if $\lambda$ is a gauge transformation of a Chern-Simons theory the Lagrangian is not invariant. The genetalization is straightforward but we refer to \cite{Tachikawa:2006sz} for a discussion of this case.} As a top form in the manifold the Lagrangian satisfies $\delta_\epsilon L=\delta_\xi L=d(\xi\inter L)$, then, $M_\epsilon =\xi\inter L$. Here and in the following we assume $\delta \xi=0$.
 Then, we have
\ba
\delta M_\epsilon & =  &\xi\inter(E\delta \Phi)+\xi\inter d\Theta(\delta\Phi).
\ea 
On the other hand
\ba
\delta_\epsilon\Theta(\delta\Phi)=\delta_\xi\Theta(\delta\Phi) = d\xi\inter \Theta(\delta\Phi)+\xi\inter d\Theta(\delta\Phi).
\ea
And, 
\ba
\Omega(\delta,\delta_\epsilon)&=&  -d\xi\inter \Theta(\delta\Phi)-\Theta([\delta,\delta_\epsilon]\Phi)+\xi\inter(E\delta\Phi)-\delta S_\epsilon +d\delta \widetilde Q_\epsilon,
\label{sympstruct}
\ea
after explicitly using the equations of motion and the linearized equations of motion, we obtain a simple expression for the presymplectic structure density
\ba
\Omega(\delta,\delta_\epsilon)  & \widehat{\widehat = }& d\left(\delta \widetilde Q_\epsilon-\xi\inter \Theta(\delta\Phi)\right)-\Theta([\delta,\delta_\epsilon]\Phi).
\label{hola}
\ea

In the case the parameters $\delta\epsilon\neq 0$ are extended non-trivially on the phase space, the last term does not vanish (we still assume $\delta\xi=0$ but $\delta\lambda\neq 0$). In the examples, it is going to be the case when gauge parameters get fixed to encode exact symmetries. Analogous to the decomposition $E\delta_\epsilon\Phi=dS_\epsilon-N_\epsilon$, the term $\Theta([\delta,\delta_\epsilon]\Phi)$ can be decomposed as
\be
\Theta([\delta,\delta_\epsilon]\Phi)=dB_{\delta\epsilon}+C_{\delta\epsilon},
\label{Ccero}
\ee
such that in  $C_{\delta\epsilon}$, the varied parameters appear as factors. A similar argument that the one used to prove the off-shell Noether identity, $N_\epsilon=0$,  proves that on-shell $C_{\delta\epsilon}\ {\widehat =}\ 0$.\footnote{In (\ref{hola}) the gauge parameters $\epsilon$ and $\delta\epsilon$ are disentangled, then fixing $\epsilon=0$ and integrating over any $(D-1)$-surface the arbitrariness of $\delta\epsilon$ implies $C_{\delta\epsilon}=0$.}

We define a $(D-2)$-form in spacetime and first variation in phase space by
\ba
k_\epsilon &\equiv &\delta\widetilde  Q_\epsilon-\xi\inter\Theta(\delta\Phi)-B_{\delta\epsilon}.
\label{SC}
\ea
As an abuse of name we may refer to this quantity as the would be {\it surface charge integrand}.

In the case $\epsilon$ represents a gauge symmetry we can choose $\delta\epsilon=0$ such that $\Theta([\delta,\delta_\epsilon]\Phi)=0$. Then, equation (\ref{hola}) tells us the standard result: The presymplectic structure density for a gauge transformation is trivial, i.e., it is an exact form in spacetime\footnote{The presymplectic structure for an action $S[\Phi]=\int_{\cal M}L$ is obtained by $$(\delta_1\delta_2-\delta_2\delta_1)S[\Phi] \ \widehat{\widehat =}\ \int_{\partial\cal M} \left[\Omega(\delta_1,\delta_2)+\Theta([\delta_1,\delta_2]\Phi)\right],$$ therefore, it is defined up to an exact form in spacetime.
}
\ba
\Omega(\delta,\delta_\epsilon)& \widehat{\widehat =}& dk_\epsilon.
\label{holahola}
\ea
A gauge symmetry is a degenerate direction in the presymplectic structure.  Once integrated, last expression becomes an arbitrary boundary term that in particular can be chosen to vanish.

Now, in the particular case that $\epsilon$ generates an exact symmetry the presymplectic structure density vanishes. We call {\it exact symmetry} the condition where particular parameters $\bar \epsilon$ solve the equation $\delta_{\bar \epsilon}\Phi=0$ (Killing fields for the metric for instance). Then, because the presymplectic structure density is linear in the infinitesimal transformations, we have $\Omega(\delta,\delta_{\bar\epsilon})\ \widehat{\widehat =} \ 0$.  Therefore, for exact symmetries
\ba
dk_{\bar\epsilon }& \widehat{\widehat =} & 0.
\label{kconserved}
\ea 
The establishment of this equation is the main goal of this section. This equation is a second conservation law of one degree less than $dJ_\epsilon=0$. It has a true physical meaning because it requires the use of the equations of motion besides the property $\delta_{\bar \epsilon}\Phi=0$. As commented in subsection \ref{motsurfacecharges}, $k_{\bar \epsilon}$ is a current in the context of a generalized Noether theorem \cite{Barnich:2000zw}.

Therefore, we define the {\it surface charge} by the integral
\be
\slashed\delta Q_{\bar\epsilon} \equiv \oint k_{\bar\epsilon}.
\ee
Note that it is called {\it surface} because it is naturally defined on a $(D-2)-$manifold which is a surface in four dimensions. And more important, it is called a {\it charge} because it is {\it conserved}. This happens only because the exactness of the symmetry guarantee the conservation law (\ref{kconserved}). That makes the integral independent of the closed surface where the integration is performed. 
On the other hand, one could compute non-vanishing quantities $\slashed \delta Q_\epsilon$ for gauge symmetries $\delta_\epsilon$ but these quantities are not charges.  Following the notation proposed in \cite{Barnich:2007bf} we use $\slashed{\delta}$ to denote quantities that are not necessarily integrable on the phase space. In other words, the function $Q_{\bar \epsilon}$ such that its variation on the phase space satisfies $\delta Q_{\bar\epsilon}=\slashed{\delta}Q_{\bar \epsilon}$ may not exist. A sufficient condition for its existence is $\delta(\slashed \delta Q_{\bar\epsilon})=0$ this is the condition of integrability for the surface charge to become a finite charge.

As explained before, a gauge symmetry produces a trivial Noether current in the sense that it is conserved even off-shell (\ref{dJ=0}). Then, to use a physical symmetry or the equations of motion or nothing, in order to prove the conservation of $J_\epsilon$ does not make any difference. However, if for certain choice of the gauge parameters $\bar \epsilon$ the gauge symmetry can be made exact, it will produce a second necessarily on-shell conservation law for $k_{\bar \epsilon}$. Let us remark that choosing the gauge parameters means that we are not dealing with a gauge symmetry anymore. However, in the derivation of $dk_{\bar\epsilon }\ \widehat{\widehat =}\  0$ we make intensive use of the presence of a gauge symmetry.\footnote{This is exacly what happens in the Hamiltonian approach for asymptotic charges: The parameters generating gauge symmetry are boiled down to exact symmetries in the asymptotic region.}

In the four dimensional examples worked out in subsection \ref{bhexample}, $k_{\bar \epsilon}$ is a closed two-form in spacetime that can be used to relate quantities defined on two arbitrary disconnected closed two-surfaces which are the boundaries of a given three-volume. The integration of $k_{\bar\epsilon}$ on a closed two-surfaces is trivial if the surface is contractible to a point. In the black hole example, $k_\epsilon$ will be integrated over spheres enclosing the singularity. Note that, as we are strongly using differentiability of fields, this should be guaranteed in the three-volume as well as in its boundary. Bulk singularities and spikes in boundaries have to be treated carefully.

Two remarks regarding possible ambiguities are in order.  First, note that there is an ambiguity in the definition of $\Theta(\delta\Phi)\to\Theta(\delta\Phi) +dY(\delta\Phi)$, which percolates to an arbitrary exact form in the presymplectic structure density. However, for exact symmetries it simply vanishes and does not have any effect in the definition of $k_{\bar\epsilon}$. Second, another ambiguity could arise because $k_{\bar\epsilon}\to k_{\bar\epsilon}+d\alpha$ does not change the equation $dk_{\bar\epsilon}\  \widehat{\widehat =}\ 0$. This ambiguity is harmless as far as $k_\epsilon$ is used only integrated over closed surfaces.

\section{General Relativity}
\label{generalrelativity}

In this section we consider the action for gravity in four dimensions in the first order formalism. This formalism is fundamental in the sense that it is suitable to include fermionic fields. At the same time the metricity and parallelism properties of spacetime can be easily disentangled \cite{Zanelli:provisory}.

The language of forms allows us to write variables without doing explicit references to coordinates. We consider as independent variables the tetrad and the Lorentz connection,  $(e^I,\omega^{IJ})$, both are one-forms. The curvature two-form read $R^{IJ}=d\omega^{IJ}+\omega\indices{^I_K}\wedge \omega^{KJ}$. Besides the standard Einstein-Hilbert term we consider a cosmological constant and a topological Euler term, all them arranged in the well-known McDowell-Mansouri action \cite{MM,Aros:1999id,Wise:2006sm}. In the following we will suppress the indexes and the wedge product to make the notation compact when possible.

Therefore, the action for gravity simple reads
\be
S[e,\omega]=\kappa\int_{\cal M} \bar R \star \bar R,
\ee
with the barred curvature given by
\be
\bar R^{IJ}\equiv R^{IJ}\pm \frac{1}{\ell^2}e^I e^J,
\ee
note that as before the wedge product between forms is understood. The $\star$ stands for the dual of the internal group, in this case the Lorentz group, for instance  $\star \bar R^{IJ}=\frac{1}{2}\varepsilon\indices{^{IJ}_{KL}}\bar R^{KL}$.  The $\pm$ stands for the both possible signs of the cosmological constant. The treatment is the same then we consider both at once. The overall constant $\kappa$ has not effect in the following but we fix it to $\kappa=\pm \frac{\ell^2}{32\pi G}$ to make contact with standard approaches. We also choose the units to set the Newton constant $G=1$.

The dependence on the cosmological constant can be consistently removed at the end of the calculation by considering the limit $\ell\to \infty$. Note that the Euler term is multiplied by $\ell^2$, thus the limit can not be taken at this stage. In fact the Euler term can be thought as providing a regulator for the Einstein-Hilbert plus cosmological constant action and for the {\it finite} Noether charges derived from it \cite{Aros:1999id}. However, as far as we consider exact symmetries the present quasilocal approach is insensitive to it. This is made explicit at the end of this section.

The variation of the Lagrangian is
\be
\delta L = E_e\delta e +E_\omega \delta \omega + d\Theta,
\ee
if we get rid of the term $\Theta$ by imposing boundary condition, as we will discuss in a moment, the variational principle implies the equations of motion (putting back the indexes, $E_e\to E_I$ and $E_\omega\to E_{IJ}$)
\ba
E_I&=&\mp \frac{2\kappa}{\ell^2}\varepsilon_{IJKL}e^J \bar R^{KL}=0,
\label{EI}
\\
E_{IJ}&=&-\kappa\varepsilon_{IJKL}d_\omega\bar R^{KL}=\mp \frac{2\kappa}{\ell^2}\varepsilon_{IJKL}d_\omega e^K e^L=0,
\ea
where we use $d_\omega$ to denote the covariant exterior derivative, for instance the Bianchi identity reads $d_\omega R^{IJ}=dR^{IJ}+\omega\indices{^I_K} R^{KJ}-\omega\indices{^J_K}R^{IK}=0$, and the torsion $T^I\equiv d_\omega e^I=de^I+\omega\indices{^I_J}e^J$. The second equation is equivalent to setting the torsion equal to zero, $d_\omega e^I=0$. Because this is an algebraic equation for the Lorentz connection, $\omega$, it can be solved in terms of the tetrad, $\omega(e)$. The replacement of $\omega(e)$ in the first equation produces the usual Einstein equation with cosmological constant written in forms.

As suggested before, to have a well-posed variational principle the term
\be
\Theta=2\kappa\delta\omega\star\bar R,
\label{boundaryGR}
\ee
must vanish at the boundary of the spacetime $\cal{M}$. If we allow for an arbitrary $\delta \omega$, the following boundary condition is required\footnote{Note that the imposed boundary condition has a symmetry larger than the local Lorentz group. It is invariant under the (anti-)de Sitter group  [which contains the Lorentz group $SO(3,1)]$.

Note also that for asymptotically flat spacetimes a well-defined action principle needs a boundary term different than the Euler one. In \cite{Ashtekar:2008jw} it is shown that $\omega^{IJ}\star(e_Ie_J)$ corresponds to the Hawking-Gibbons term in the first order formalism, and therefore it allows a well-defined variational principle as well as asymptotic Hamiltonians. However, as explained in Subsection \ref{topo} boundary terms in the action do not contribute in surface charges approach that we follow.

}
\be
\left.\Theta\right|_{\partial\cal  M}=0\quad \rightarrow \quad \left.\bar R\right|_{\partial\cal  M}=0.
\label{barF=0}
\ee
Note that this condition requires an a priori knowledge of the boundary of the spacetime. In other words, we are reducing the space of solution such that the previous equation can be satisfied. The family of spacetimes with this property are named locally asymptotic (anti-)de Sitter spacetimes. On the other hand, the standard assumption $\left.\delta\omega\right|_{\partial {\cal M}}=0$ is more relaxed because it can be applied in principle to any patch of the spacetime.  However, it is a strong condition because $\omega$ is a connection and we would need to fix the gauge in the boundary too.

The approach we follow to define the surface charges is quasilocal. It is insensitive to the chosen prescription for the boundary term. The only requirement is that a well-posed variational principle exists in order to obtain the equations of motion.

The gauge symmetries of the action are general diffeomorphisms and local Lorentz transformations. The infinitesimal transformations of the fields by the local Lorentz group is
\ba
\delta_\lambda e^I&=&\lambda\indices{^I_J} e^J \\
 \delta_\lambda \omega\indices{^I_J} &=&-(d_\omega \lambda)\indices{^I_J}= -d\lambda\indices{^I_J}-\omega\indices{^I_K}\lambda\indices{^K_J}+\omega\indices{_J^K}\lambda\indices{^I_K},
\ea
where $\lambda^{IJ}=-\lambda^{JI}$ are the parameters of the infinitesimal Lorentz transformation $\Lambda\indices{^I_J}\approx \delta\indices{^I_J}+\lambda\indices{^I_J}$. Remember that it is a gauge symmetry, that is, the group elements take different values at different points of the manifold.

The infinitesimal transformations of the fields due to diffeomorphisms are normally assumed to be generated by an arbitrary vector field $\xi$ through a Lie derivative 
\ba
\tilde\delta_\xi e&=&\Lie_\xi e = d(\xi\inter e)+\xi \inter (de) 
\label{Lies1}\\
\tilde\delta_\xi \omega&=&\Lie_\xi \omega = d(\xi\inter \omega)+\xi \inter (d\omega),
\label{Lies2}
\ea
where in the second equality we use the Cartan formula. However, note that due to the presence of exterior derivatives they are not homogeneous under local Lorentz transformation. The intuitive interpretation of $\delta_\xi e$ and $\delta_\xi \omega$ as infinitesimal variation require them to be homogeneous under the action of the local Lorentz group. More precisely, if we attach ourselves to the intuitive idea of {\it variations as comparison of fields in a neighbourhood}, $\delta e \approx e'-e$, we expect them to have a covariant transformation under the local Lorentz group.  This criteria is not satisfied by the infinitesimal diffeomorphism transformation presented before, and therefore we correct (\ref{Lies1})-(\ref{Lies2}) by eliminating the non-homogeneous part. This can be done by adding an infinitesimal Lorentz transformation with a parameter $\xi\inter \omega$. For a recent discussion see \cite{Jacobson:2015uqa}. This corrects the non-homogeneous part of both transformations at once, and we get
\ba
\delta_\xi e&=& \Lie_\xi e+\delta_{\xi\inter\omega}e= d_\omega(\xi\inter e)+\xi \inter (d_\omega e) 
\label{Lies1c}\\
\delta_\xi \omega&=&\Lie_\xi \omega+\delta_{\xi\inter\omega}\omega= \xi \inter R.
\label{Lies2c}
\ea
 Another way to think about this, is that in the transformation of the tetrad, the exterior derivative $d$ is promoted to a covariant exterior derivative $d_\omega$, while in the transformation of the Lorentz connection, because of the identity $d(\xi\inter \omega)+\xi \inter d\omega=d_\omega(\xi\inter \omega)+\xi \inter R$, the ill-transforming part, $d_\omega(\xi\inter \omega)$, is subtracted.

Therefore, the general infinitesimal gauge transformations, involving diffeomorphisms and local Lorentz transformations, with parameters $\epsilon=(\xi,\lambda)$, which are themselves homogeneous, are\footnote{Still other prescriptions for the infinitesimal transformations are possible. For instance the ones recently introduced in \cite{Montesinos:2017epa} differ from (\ref{transftetrad}) and (\ref{transfomega}) by a term depending on the equation of motion. These kind of terms are known as {\it trivial symmetries} because they are present in any theory (section 3.1.5 in \cite{Henneaux:1992ig}).}
\ba
\delta_\epsilon e &=&\Lie_\xi e+\delta_{(\xi\inter\omega+\lambda)}e= d_\omega (\xi\inter e)+\xi\inter (d_\omega e)+\lambda e
\label{transftetrad}\\
\delta_\epsilon \omega &=&\Lie_\xi \omega+\delta_{(\xi\inter\omega+\lambda)}\omega =\xi\inter R-d_\omega\lambda. 
\label{transfomega}
\ea

Now, we follow the procedure detailed in section \ref{surfacecharges} to obtain the surface charges for General Relativity. The Lagrangian transforms as a total derivative, $
\delta_\epsilon L = \Lie_\xi L =d(\xi\inter L)$. Under a symmetry transformation
\be
d(\xi\inter L-\Theta(\delta_\epsilon\omega))=E_e\delta_\epsilon e +E_\omega \delta_\epsilon \omega ,
\ee
using explicitly the symmetry transformation on the variables we can mimic (\ref{noether0})
\ba
E_e\delta_\epsilon e +E_\omega \delta_\epsilon \omega &=&d\left[E_\omega \lambda-E_e\xi\inter e\right]+\left(E_e e-d_\omega E_\omega \right)\lambda+d_\omega E_e\xi\inter e+E_e\xi\inter d_\omega e+E_\omega \xi\inter R\quad\quad\ \ \\
&=&d\left[E_\omega \lambda-E_e\xi\inter e\right].
\ea
In the first line we used exact forms to have the gauge symmetry parameters either inside an exact form or as a factor of a term. In the second line we used the Noether identities: $d_\omega E_\omega-E_e e =0$ and $d_\omega E_e\xi\inter e+E_e\xi\inter d_\omega e+E_\omega\xi\inter R=0$, which are a rewriting of the standard Bianchi identity. Then, we define
\be
J_\epsilon\equiv \Theta(\delta_\epsilon\omega)-\xi\inter L-E_e\xi\inter e+E_\omega \lambda,
\label{JGR0}
\ee
that trivially satisfies $d J_\epsilon=0$. Explicit computation results in an exact three-form that depends just on the gauge parameter of the Lorentz symmetry $\lambda$
\be
 J_\epsilon=J_\lambda= -\kappa\ d\left(2\lambda\star \bar R\right)
\label{JGR}
\ee
In spite its trivial conservation one may try to use this current to define global charges associated to exact symmetries at the asymptotic boundary. The result is non-trivial as shown in \cite{Aros:1999id}. It is in this context that the Euler density becomes crucial, to accomplish the boundary condition $\left.\bar R\right|_{\partial{\cal M}}=0$. This regularize the symplectic structure such that there is no leaking on the boundary for locally asymptotic (anti-)de Sitter spacetimes. Then, global finite charges can be asymptotically computed through that method.\footnote{Note that in \cite{Aros:1999id} the integrand to define the charge is $\xi\inter \omega_{IJ} \star \bar R^{IJ}$, which is gauge dependent, then, an explicit gauge fixing at the boundary is required such that it also respects the gauge dependent asymptotic symmetry condition $\left.\Lie_\xi e^I\right|_{\partial{\cal M}}=0$. That is equivalent to our expression where we can use the integrand $\lambda_{IJ}\star \bar R^{IJ}$ and can fix $\lambda^{IJ}$ by the exactness condition expressed below in (\ref{lambda2}). However, these expressions are explicitly Lorentz invariant.}

But here we intend for a quasilocal definition of charges. Following the prescription of section \ref{surfacecharges}, we make a step further and perform an arbitrary variation of $J_\lambda$ and compute each term of (\ref{sympstruct}). To compute the $B_{\delta\epsilon}$ contribution note that 
\ba
\Theta([\delta,\delta_\epsilon])&=&2\kappa[\delta,\delta_\epsilon]\omega\star\bar R\n\\
&=&-2\kappa\ d_\omega(\delta\lambda+\xi\inter\delta\omega)\star \bar R\n\\
&=&-2\kappa\ d[(\delta\lambda+\xi\inter\delta\omega)\star \bar R]+2\kappa (\delta\lambda+\xi\inter\delta\omega)\star d_\omega \bar R,
\label{BGR}
\ea
where we used $[\delta,\delta_\epsilon]\omega=\delta_{(\delta\lambda+\xi\inter\delta\omega)}\omega$. As it was shown in general, (\ref{Ccero}), the last term in the third line, correponding to $C_{\delta\epsilon}$, vanishes on-shell.

Then, from (\ref{boundaryGR}), (\ref{JGR}), and (\ref{BGR}) we obtain the surface charge integrand for General Relativity
\ba
k_\epsilon &=&-2\kappa \left(\lambda\star \delta \bar R-\delta\omega\star\xi\inter\bar R\right).
\label{kGR}
\ea
Now, if the exact symmetry condition is satisfied, we have $dk_{\bar\epsilon}=0$, and we can define surface charges $\slashed \delta Q_{\bar\epsilon}=\oint k_{\bar \epsilon}$. In the example we will be able to integrate the varied quantities on the phase space to find $Q_{\bar\epsilon}$. The charges are varied on the phase space, through a family of solutions. The study of the phase space for a family of solutions can be done explicitly, for instance, when considering the variation of the integration constants that appear in a solution.

For completeness we write down the presymplectic structure density for pure gravity
\ba
\Omega(\delta_1,\delta_2)&\widehat{\widehat =}&2\kappa\left(\delta_2\omega\star \delta_1 \bar R-\delta_1\omega\star \delta_2 \bar R\right)\\
&\widehat{\widehat =}&-\frac{1}{8\pi}\delta_{[1}\omega^{IJ}\ \delta_{2]}\Sigma_{IJ}+2\kappa d(\delta_1\omega^{IJ}\star\delta_2\omega_{IJ}),
\ea
where we used that $\delta R =d_\omega\delta\omega$, the value of $\kappa$, and the definition $\Sigma_{IJ}\equiv\frac{1}{2}\varepsilon_{IJKL}e^K\wedge e^L$. The first term is the conjugate pair of gravity variables $(\omega^{IJ}, \Sigma_{IJ})$. The second term, consequence of the Euler term in the action, is an exact form and disappears when the density is integrated on a smooth boundary of a manifold, $\partial {\cal M}$. Similarly, the Euler contribution to the surface charge will not have any effect because for exact symmetries it becomes and exact form. Explicitly, the contribution of the Euler term to (\ref{kGR}) is
\be
k_\epsilon^{Euler}=-2\kappa\left[\lambda\star \delta R-\delta\omega\star\xi\inter R\right]=-2\kappa\left[d(\lambda\star \delta\omega)+\delta_\epsilon\omega\star\delta\omega\right].
\label{keuler}
\ee
This confirms that the procedure is not affected by the action boundary terms.

Remember, to guarantee that $k_\epsilon$ is closed we need an exact symmetry such that $\Omega(\delta,\delta_\epsilon)\ \widehat{\widehat =}\ 0$. Therefore, we have to solve the parameters $\epsilon=(\xi,\lambda)$ such that
\ba
\delta_\epsilon e&=&0,
\label{deltae}\\
\delta_\epsilon\omega &=&0.
\ea
In the following $\epsilon=(\xi,\lambda)$ are solutions of the previous equation. The condition $\delta_\epsilon e=0$ imposes a general relation between $\lambda$ and $\xi$. Exact symmetries are on-shell, thus we use $d_\omega e=0$, and solve $\lambda$ from (\ref{deltae})
\be
\lambda^{IJ}=e^I\inter d_{\omega}(\xi\inter e^J)=e^{I\mu}e^{J\nu}\nabla_{[\mu}\xi_{\nu]},
\label{lambda2}
\ee
where $e^I\inter$ is the interior product such that $e^I\inter e_J=\delta^I_J$, in coordinates components it is $e^{I\mu}\partial_\mu\inter$. In the second equality we exhibit the solution in components with $\nabla_\mu$ the spacetime covariant derivatve. This relation is a sufficient condition that gauge parameters should accomplish to encode an exact symmetry. Note that the Killing equation, $\Lie_\xi g=0$ with the metric $g=e_I\otimes e^I$, is a direct consequence of $\delta_\epsilon e=0$. The Killing equation in coordinate components, $\nabla_{(\mu}\xi_{\nu)}=0$, can also be seen directly in the rightest expression for $\lambda$: It is encoded in the fact that $\lambda^{IJ}$ is antisymmetric. On the other hand, because $\omega=\omega(e)$, the condition $\delta_\epsilon\omega=0$ holds trivially.

Therefore, we have obtained the expresion of for the surface charges in the tetradic first order formalism (integration of (\ref{kGR})). We have also shown that the exact symmetries condition for the tetrad is the Killing equation in this language.

As a final remark, note that there is a second and straight way to obtain the same result. Consider the following expression for the {\it contracting homotopy operator} discussed in the appendix \ref{appendix}
\be
I_{\delta e,\delta \omega}\equiv \delta e^I\wedge \frac{\partial\ \ \ }{\partial d_\omega e^I}+\delta\omega^{IJ}\wedge\frac{\partial\ \ \ }{\partial R^{IJ}}.
\ee
Acting with this operator on $S_\epsilon$ we obtain the following surface charge integrand
\be
k'_\epsilon\equiv I_{\delta e,\delta \omega} S_\epsilon=k_\epsilon-k_\epsilon^{Euler}.
\label{homotopy}
\ee
It has the advantage of giving directly the term that does not have a contribution from the action boundary term. Note that in the general framework $S_\epsilon $ is the boundary term independent part of $J_\epsilon$. However, the difference is harmless because for exact symmetries we have $\oint k'_\epsilon=\oint k_\epsilon$, and therefore both prescriptions are equivalent. This fact has a straightforward generalization for Lovelock theories in $D$ dimensions, section \ref{dworld}.

\subsection{Topological and boundary terms effect on surface charges}
\label{topo}
Surface charges are insensitive to action boundary terms. This is a remarkable property that is in high contrast with usual Noether procedures to compute charges (see for instance the recent review \cite{Corichi:2016zac}). To see how this happens let us add a boundary term to the Lagrangian: $L\rightarrow L+d\alpha$, with the assumption that $\alpha$ is gauge invariant, $\delta_\epsilon \alpha =\Lie_\xi \alpha$. Now we can repeat the procedure of section \ref{surfacecharges} by  keeping track of this boundary term. We have $J_\epsilon \rightarrow J_\epsilon + d(\xi \inter \alpha)$, and the surface charge potential
\begin{equation}
k_\epsilon \rightarrow k_\epsilon + \delta(\xi\inter \alpha) - \xi \inter \delta \alpha=k_\epsilon,
\end{equation}  
where we used $\delta\xi=0$. Thus, there is not change at all for the surface charges.

The previous case is quite general. For example in $D=4$ the Nieh-Yan topological term, with density $d(e^IT_I)=T^IT_I-e_Ie_JR^IJ$, is in this category. However, there are examples where $\alpha$ is not gauge invariant. This is the case for the Euler or the Pontryagin topological terms, with densities $R^{IJ}\star R_{IJ}$ and $R^{IJ}R_{IJ}$, respectively. Both are exterior derivatives of Chern-Simons Lagrangian which are gauge quasi-invariant forms \cite{Zanelli:provisory}. For the Euler term the expression (\ref{keuler}) shows that it does not affect the surface charge. A similar computation for the Pontryagin yields
\begin{equation}
k_\epsilon^{Pontryagin}=-2\kappa\left[\lambda \delta R-\delta\omega\xi\inter R\right]=-2\kappa\left[d(\lambda \delta\omega)+\delta_\epsilon\omega\delta\omega\right],
\label{pontryagin}
\end{equation}
then, surface charges are blind to the Pontryagin topological term too.

Finally, in $D=4$ we may also be interested in using the Holst term density, $e_Ie_JR^{IJ}$, inside the gravity action. This is not a topological term by itself but a part of the Nieh-Yan, and it does not affects the equations of motion either. To deal with it note that $e_Ie_JR^{IJ}=T^IT_I-d(e^IT_I)$. The second term was already studied, then, it is enough to keep track of $T^IT_I$ in the computation of surface charges potential. This term also does not produce any changes because already at the level of the presymplectic structure density, $\Omega(\delta,\delta_\epsilon)$, the contributions are all proportional to the torsion $T$ and therefore vanish on-shell.

Then, boundary terms, and in particular topological terms, do not affect the surface charges. Note that this is already explicit for surface charges computed through the contracting homotopy operator (\ref{homotopy}) because it depends only of $S_{\bar\epsilon}$ and not of the Lagrangian. In this sense here we have stressed what is already indirectly known due to the fact that surface charges obtained through both methods are equivalent (appendix \ref{appendix}).

\section{Electromagnetism}
\label{electro}

Before dealing with the more general case of General Relativity coupled to Electromagnetism we briefly review the pure electromagnetic theory. Because diffeomorphisms are not a gauge symmetry here, the procedure is simpler. The variable is the connection one-form $A=A_\mu dx^\mu$. The field strength two-form is $F=dA$, and it posses the $U(1)$ gauge symmetry, $A\to A+d\Lambda$. 

The action is 
\be
S[A]=\alpha\int_{\cal M} F * F,
\ee
with $\alpha=-1/8\pi$, and where the Hodge dual $*$ acting on the field strength in coordinate components or in tetrad components is respectively $* F=\frac{e}{2!}\varepsilon_{\mu\nu\alpha\beta}F^{\alpha\beta}dx^\mu\wedge dx^\nu=\frac{1}{2!}\varepsilon_{IJKL}F^{IJ}e^K\wedge e^L$, with $e=\det(e\indices{^I_\mu})$ and $F^{IJ}=e^{I\mu}e^{J\nu}F_{\mu\nu}$.

The variation of the Lagrangian is
\ba
\delta L(A) &=& E_A \delta A+d\Theta(\delta A)\\
&=&-2\alpha\left(d* F \right) \delta A+d\left(2\alpha\delta A* F\right),
\label{varEM}
\ea
where $\delta *=*\delta$ because $e^I$ is not a dynamical field for this theory. As before, the boundary term should vanish. An option is to fix the connection at the boundary and consequently the gauge symmetry. Another option that does not restrict the connection is to assume a vanishing field strength at the boundary, $\left.F\right|_{\partial \cal M}=0$. The infinitesimal symmetry gauge transformation is $\delta_\lambda A=-d\lambda$, and applying it to the variation of the Lagrangian, $E_A(-d\lambda)=d(E_A\lambda)-dE_A \lambda$ we get the  Noether identity, which is the trivial equation $-dE_A=2\alpha d(d*F)=0$. The Lagrangian is invariant under the gauge transformation, therefore, we have
\ba
J_\lambda=E_A\lambda+\Theta(\delta A)=-2\alpha d(\lambda * F).
\label{currentEM}
\ea
Note the similarity with the corresponding equation for General Relativity (\ref{JGR}).

The presymplectic structure density is $\Omega(\delta_1,\delta_2)\ \widehat{\widehat =}\ -4\alpha \delta_{[1}A*d\delta_{2]}A$. And for a gauge symmetry, $\Omega(\delta,\delta_\lambda)\ \widehat{\widehat =}\ dk_\lambda$, we get the surface charge integrand
\be
k_\lambda=-2\alpha \lambda  *\delta F.
\label{kEM}
\ee
The exact symmetry condition $\delta_\lambda A=0$ is solved for $\lambda=\lambda_0=constant$, such that the gauge symmetry turns into a rigid symmetry. Note that here the exact condition is independent of the fields and admits a general solution. Then, we have $dk_{\lambda_{0}}=0$ which can be integrated in a three-surface $\Sigma$ enclosed by a two-surface $S$, to define
\ba
\slashed\delta Q_{\lambda_0}=\frac{\lambda_0}{4\pi}\oint_S *\delta F,
\ea
where we have restored the value of $\alpha$. For simplicity the parameter $\lambda_0$ is chosen to be a constant in the phase space. Then, the variation can be trivially removed by an integration on phase space. We set the integration constant to zero. Then, we obtain the definition of the electric charge, enclosed by the surface $S$
\be
Q_{\lambda_0}=\frac{\lambda_0}{4\pi}\oint * F,
\ee
the conservation $dk_{\lambda_0}=0$ ensures that for any other surface, $S'$ obtained by a continuous deformation of $S$, the electric charge is the same. If there are not sources $S$ can be contracted to a point and all charges are zero. This finishes the analysis of surface charges for Electromagnetism.

For completeness we derive the Noether current for a background spacetimes with rigid symmetries. Note that the spacetime may not be a solution of the Einstein equation. The rigid symmetries are controlled by a Killing field $\xi$ such that $\delta_\epsilon e^I=0$, where $\epsilon=(\xi,\lambda^{IJ})$ and $\lambda^{IJ}$ is given by (\ref{lambda2}). The connection $A$ suffers from the same ambiguity of any gauge variable, then a Lie derivative on it, $\Lie_\xi A=\xi\inter F+d\xi\inter A$ may be changed by an arbitrary gauge transformation while keeping the same information. Repeating the argument that led us to (\ref{Lies2c}), the symmetry infinitesimal transformation which is also gauge invariant is 
\ba
\delta_\xi A=\xi\inter F,
\label{transA}
\ea 
this transformation sometimes is called `improved.' Note that while being assumed as a physical transformation it is not an exact symmetry transformation. Applied to the Lagrangian $\delta_\xi L=d(\xi\inter L)$. Because $\xi$ is not an arbitrary parameter, but a fixed Killing field, there is not a Noether identity associated to it. The current $J_\xi\equiv\Theta(\delta_\xi A)-\xi\inter L$, is a true Noether current because it is conserved just on-shell
\ba
d(\Theta(\delta_\xi A)-\xi\inter L)&=&-E_A\delta_\xi A\ {\widehat =}\ 0,
\ea 
explicitly
\ba
J_\xi=\alpha\left(\xi\inter F * F-F\xi\inter * F\right),
\ea
which is the dual of the standard four-current of Electromagnetism written in forms, $*J_\xi=j=j_\nu dx^\nu=\xi^\mu T^{{\text{\tiny EM}}}_{\mu\nu}dx^\nu$, with $T^{{\text{\tiny EM}}}_{\mu\nu}$ the electromagnetic stress-energy tensor. Note that again $\delta_\xi A\neq \Lie_\xi A$. Instead, we used an infinitesimal transformation that is covariant (actually invariant) under the gauge transformation. This subtelty has produced debate in the literature and the transformation (\ref{transA}) has been settled as the right one because it produces a gauge invariant current (see for instance section 2.3 in the recent review \cite{Banados:2016zim}). However, the Noether current belongs to an equivalence class of currents related by exact forms (and possibly terms vanishing on-shell). The gauge ambiguity in the chosen transformation for the symmetry (\ref{transA}) just change the representative current of the class. To define a charge the equation, $d J_\xi=0$, should be integrated on a four-dimensional manifold, which is splittable in two pieces. This procedure is insensitive to the ambiguity, and therefore the Noether charge. 

\section{General Relativity and Electromagnetism}
\label{generalelectro}

Using the results of the previous sections the extension to the coupled theory is easy. Here we use $*$ for the Hodge dual and $\star$ for the group dual. We also make explicit some indexes to differentiate the $U(1)$ infinitesimal gauge parameter $\lambda$, from the $SO(3,1)$ infinitesimal gauge parameter $\lambda^{IJ}$.

The Lagrangian and its variation are given by
\ba
L&=&\kappa \bar R_{IJ}\star \bar R^{IJ}+ \alpha F * F\\
\delta L &=& E'_e\delta e+E_\omega \delta \omega +E_A\delta A+d\Theta(\delta_\omega,\delta A) ,
\ea
where $E'_e\delta e= \left(E_I+M_I\right)\delta e^I$. We use the notation $e_I\inter =e\indices{_I^\mu}\partial_\mu\inter$. $E'_e$ is the sum of  the equation of motion of pure gravity (\ref{EI}) plus the contribution due to the electromagnetic stress tensor written as a form $M_I\equiv \alpha(e_I\inter F * F-Fe_I\inter * F)\sim \alpha e\indices{_I^\mu}T^{EM}_{\mu\nu}*dx^\nu$. The boundary term reads
\ba
\Theta(\delta \omega,\delta A)=2\kappa \delta \omega_{IJ} \star \bar R^{IJ}+2\alpha \delta A * F.
\ea
The infinitesimal gauge symmetry transformations are controlled by the parameters $\epsilon=(\xi,\lambda^{IJ},\lambda)$ corresponding to diffeomorphisms, Lorentz local symmetry, and $U(1)$ local symmetry, respectively. The transformations are the same for the gravity fields, (\ref{transftetrad}) and (\ref{transfomega}). For the electromagnetic field we need to consider that it also transforms by diffeomorphisms. As discussed in the previous section the improved version is $\delta_\epsilon A=\xi\inter F-d\lambda$.

Following the procedure of section \ref{surfacecharges} we obtain
\be
J_\epsilon=-2\ d\left(\kappa \lambda_{IJ}\star \bar R^{IJ}+\alpha \lambda * F\right),
\ee
that is simply the sum of gravity and electromagnetic off-shell constributions found previously, (\ref{JGR}) and (\ref{currentEM}). While expected this is non-trivial because there is a non obvious off-shell cancellation among the terms proportional to $\xi$ appearing in $\Theta(\delta_\epsilon \omega,\delta_\epsilon A)$, $\xi\inter L$, and $S_\epsilon$.

The full surface charge integrand is
\ba
k_\epsilon=-2\kappa\left(\lambda_{IJ}\star \delta\bar R^{IJ}-\delta \omega_{IJ} \star \xi\inter\bar R^{IJ}\right)-2\alpha\left(\lambda \ \delta * F- \delta A\ \xi\inter* F\right).
\label{kGREM}
\ea
This is the sum of equations (\ref{kGR}) and (\ref{kEM}) plus a contribution due to diffeomorphism transformation of the electomagnetic field.
Now, to ensure that $dk_\epsilon=0$ we need the exactness of the symmetries $\delta_\epsilon e=0$, which is already solved by (\ref{lambda2}), but we also need
\be
\delta_\epsilon A=\xi\inter F-d\lambda=0,
\label{exactA}
\ee
from this equation $\lambda$ can be solved in general. In coordinates components it is equivalent to solve $\lambda$ from $\frac{1}{2}\xi^\mu F_{\mu\nu}=\partial_\nu \lambda$. It is in fact the equation for the electric potential for an electromagnetic field projected with $\xi$. We note that $\lambda_0=constant$  is solution of the homogeneous equation, therefore, if $\bar \lambda$ is solution of the inhomogeneous one, we have $\lambda=\bar \lambda+\lambda_0$. The $\lambda_0$ plays exactly the same r\^ole that in pure Electromagnetism and therefore implies the conservation of the electric charge. 

Note that in the surface charges formalism the definition of the electric charge and the charges due to spacetime Killing symmetries are on the same foot.

Before discussing an example let us remark the linearity property of the surface charge integrands. In the general derivation of section \ref{surfacecharges} we used the assumption $\delta\xi=0$. However, the obtained formula (\ref{kGREM}) is explicitly linear in the vector field generating diffeomorphism and in all the gauge parameters, i.e.
\be
\alpha_1 k_{\epsilon_1}+\alpha_2 k_{\epsilon_2}=k_{\alpha_1\epsilon_1+\alpha_2\epsilon_2},
\label{linearity}
\ee
where $\alpha_{1,2}$ can be {\it arbitrary functions} on the phase space. Thus, if $k_{\epsilon_1}$ and $k_{\epsilon_2}$ are closed forms for exact symmetries generated by $\epsilon_1$ and $\epsilon_2$, then $k_{\epsilon_3}=k_{\alpha_1\epsilon_1+\alpha_2\epsilon_2}$ is also a closed form for the exact symmetry generated by $\epsilon_3$ with the precise identification $\epsilon_3=\alpha_1\epsilon_1+\alpha_2\epsilon_2$.  This fact is exploited in what follows.

\subsection{Charged and rotating black hole}
\label{bhexample}

As an example we apply the result to a particular black holes familiy. We show that surface charges are compatible with the ones obtained through the standard asymptotic analysis. Then, we show how the quasilocal nature of surface charge allow to have a first law of black hole mechanics without relying on the asymptotic structure of spacetime \cite{Barnich:2003xg}. Note that this quasilocal perspective is the best that can be done when the black hole is embedded in and asymptotically de Sitter spacetime. 

We consider a black hole solution family which is electrically charged, rotating, and satisfies the asymptotically constant curvature boundary conditions (\ref{barF=0}). It is known as the (anti-)de Sitter Kerr-Newman family. A possible tetrad and electromagnetic potential describing the solution are
\ba
e^0&=&\frac{\sqrt{\Delta_r}}{\rho}\left(dt-\frac{a\sin^2\theta}{\Xi}d\phi\right),\quad e^1=\frac{\rho}{\sqrt{\Delta_r}}dr\ ,\n\\
e^2&=&-\frac{\rho}{\sqrt{\Delta_\theta}}d\theta\ ,\quad e^3=\frac{\sqrt{\Delta_\theta}\sin\theta}{\rho}\left(adt-\frac{a^2+r^2}{\Xi}d\phi\right),\\
A&=&-\frac{qr}{\rho^2}\left(dt-\frac{a\sin^2\theta}{\Xi}d\phi\right),
\ea
with $\Delta_r=(a^2+r^2)\left(1\pm\frac{r^2}{\ell^2}\right)-2mr+q^2$, $\Delta_\theta=1\mp\frac{a^2}{\ell^2}\cos^2\theta$, $\rho^2=r^2+a^2\cos^2\theta$, and $\Xi=1\mp\frac{a^2}{\ell^2}$. The upper sign is reserved for the anti-de Sitter family and the lower one for the de Sitter one. We stress that it is possible to use another set of variables related by a gauge transformation, but as the procedure is explicitly gauge invariant it will not have any impact on the results.  In particular to rotate $e^I$ by an arbitrary Lorentz transformation or to add a term of the form $d\tilde\lambda$ to $A$ has not effect. From the equation $d_\omega e^I=0$ we solve the connection and compute: $\delta\omega^{IJ}$, $\delta\bar R^{IJ}$, $\delta A$, and $\delta * F$. At this level we have reduced the phase space to the particular family solution spanned by the parameters $(m,a,q)$, thus, the variation $\delta$ acts only on functions of those parameters.

In the metric formalisms $\partial_t$ and $\partial_\phi$ are two independent Killing fields. Through the solution of the exactness conditions for $e^I$, (\ref{lambda2}), we get $\lambda^{IJ}_t$ and $\lambda^{IJ}_\phi$ respectively. Similarly through the exactness conditions on $A$, (\ref{exactA}), we obtain the corresponding $\lambda_t$ and $\lambda_\phi$. Now we have the ingredients to compute surface charges. Plugging all the quantities in (\ref{kGREM}) we get the associated integrands $k_t$ and $k_\phi$, one for each symmetry. The spacetime described by $e^I$ has non-contractible spheres due to the singularity. The integration can be performed over {\it any} two-surface enclosing the singularity. The surface charges associated to the exact symmetries generated by $\epsilon_t=(\partial_t,\lambda^{IJ}_t,\lambda_t)$ and $\epsilon_\phi=(\partial_\phi,\lambda^{IJ}_\phi,\lambda_\phi)$ are
\ba
\slashed{\delta} Q_t&=&\oint k_t=\frac{\delta m}{\Xi}\pm\frac{3am\delta a}{\ell^2\Xi^2}\\
\slashed{\delta} Q_\phi&=&\oint k_\phi=-\frac{a\delta m}{\Xi^2}+\left(\frac{3}{\Xi^2}-\frac{4}{\Xi^3}\right)m\delta a.
\ea
The exactness condition $\delta_\epsilon A=0$ has a further independent solution for a constant $\lambda_0$ such that $\delta_{\lambda_0}A=-d\lambda_0=0$. The corresponding exact symmetry parameter is $\epsilon_{\lambda_0}=(0,0,\lambda_0)$ and the surface charge is
\be
\slashed{\delta} Q_{\lambda_0}=\oint k_{\lambda_0}=\frac{\lambda_0}{4\pi}\oint\delta *F=-\lambda_0\left(\frac{\delta q}{\Xi}\pm\frac{2a q\delta a}{\ell^2\Xi^2}\right).
\ee
To proceed now we have two strategies: To fit the scheme in the results from the asymptotic picture or to insist with a quasilocal approach.  We sketch both.

{\it Asymptotic strategy:} In order to fit with the asymptotic picture we can exploit the linearity of each surface charge, (\ref{linearity}),  and to adjust the freedom of the gauge parameters in the phase space to obtain the standard integrated charges (see for instance \cite{Caldarelli:1999xj})
\ba
M&\equiv&Q_{\xi=\partial_t\mp(a/\ell^2)\partial_\phi}=\frac{m}{\Xi^2}\\
J&\equiv&Q_{-\phi}=\frac{am}{\Xi^2}\\
Q&\equiv&Q_{\lambda_0=-1}=\frac{q}{\Xi}.
\ea
The surface charge associated to $\partial_t$ is not integrable. However, the linearity property allows us to choose a different combination of the symmetry parameter $\xi\equiv \partial_t\mp\frac{a}{\ell^2}\partial_\phi$ that in fact produces an integrable charge. Note that $\xi$ is phase space dependent: $\delta \xi\neq 0$.

The charges satisfy the equation known as the black hole fundamental thermodynamics relation,
\be
M^2=\frac{S}{4\pi}\left(1\pm\frac{S}{\pi\ell^2}\right)^2+J^2\left(\frac{\pi}{S}\pm \frac{1}{\ell^2}\right)+\frac{Q^2}{2}\left(1\pm\frac{S}{\pi\ell^2}+\frac{\pi Q^2}{2S}\right),	
\ee
 which can be obtained by rewriting the condition $\Delta_r=0$ in terms of the integrated charges plus $S\equiv A/4$ with the area of the horizon $A=4\pi(r_+^2+a^2)$. The horizon is at $r=r_+$ with $r_+$ the largest solution of $\Delta_r=0$. From the last equation it follows
\be
\delta M=T\delta S+\Omega\delta J+\Phi \delta Q,
\ee
where the parametrization of the phase space is done with the integrable charges $S,\ J$, and $Q$ such that $M=M(S,J,Q)$. Then, the quantities $T$, $\Omega$, and $\Phi$ have the usual physical interpretation: $T\equiv \frac{\partial M}{\partial S}$ coincides with the Hawking temperature, $\Omega\equiv \frac{\partial M}{\partial J}$ is the horizon angular velocity, and $\Phi\equiv \frac{\partial M}{\partial Q}$ the electric potential at the horizon.

The drawback of this logic line is that it relies on previous results. Ultimately, it relies on a choice of asymptotic tailing of the field components which admits an asymptotic time symmetry and allow us to make sense of a general asymptotic mass definition. In the practice we fixed the gauge parameters to obtain a known mass expression obtained with the asymptotic method. That, for the case of anti-de Sitter certainly relies on an asymptotic analysis. However, in the cases of asymptotically de Sitter spacetimes there is no notion of time symmetry in the asymptotic region and not a physical argument to define a standard mass,\footnote{Remember that the boundary of asymptotically de Sitter spacetimes are two disconnected three-dimensional spacelike regions, one for the infinite past and one for the infinite future, and therefore none of them have a notion of time symmetry.} we just kept the $\pm$ in the formulae because it is consistent. Thus, given the quasilocal construction just developed, a pertinent question is: Is there a way to derive the first law of black hole mechanics based just on a quasilocal data?

{\it Quasilocal strategy:} To use the area of the black hole horizon as a starting point is a possibility. The area of the horizon is a well-defined quasilocal quantity which is also a finite function of the parameters of the solution. The variation of $A(m,a,q)$ on the phase space can be expressed as a combination of all the surface charges
\ba
\delta A=\oint k_\epsilon &=&\alpha(m,a,q)\slashed{\delta} Q_t+\beta(m,a,q)\slashed{\delta}Q_\phi+\gamma(m,a,q)\slashed{\delta}Q_{\lambda_0}\\
&=&\alpha'(m,a,q)\oint k_\xi+\beta'(m,a,q)\oint k_{-\phi}+\gamma'(m,a,q)\oint k_{\lambda_0=-1}\\
&=&\frac{4}{T}\delta M-\frac{4\Omega}{T}\delta J-\frac{4\Phi}{T}\delta Q,
\ea
we expressed the freedom of the gauge parameter on the phase space explicitly. On the second line we expanded in a linear combination of integrable quantities. The problem reduces to find the coefficient accompanying the integrated charges. Certainly, we already know that the result, expressed in the third line, is a rearrangement of the first law presented just before. However,  we stress the difference in the logic, in this approach the mass appears as an integrable charge computed quasilocally without the need of any asymptotic structure or physical interpretation to define it. This quantity coincides with the mass obtained by an asymptotic definition when such definition is at disposal, but it is more general because it requires just a quasilocal description of the spacetime.

Note that the two closed two-surfaces where the integration of $k_\epsilon$ is performed, besides enclosing the singularity, are arbitrary.  For a matter of physical interpretation, the one of $\oint k_\epsilon$ can be chosen to be a section of the horizon, thus being associated with the area, while for each of the other integrals it can be chosen at convenience producing for each of them the same value of the charges. This freedom plus the gauge invariance of $k_\epsilon $ can be exploited to compute the quantities easily. For instance when a bifurcated horizon is at disposal the pullback of a particular combination of the Killing fields vanishes on it and the surfaces charge formula simplifies considerably.

Summarizing, from this  second perspective the first law of black hole mechanics is a consequence of the expansion of $\delta A=\oint k_\epsilon$ into independent integrable quantities. One for each independent exact symmetry $\epsilon_i$. To accomplish integrability the symmetry parameters should satisfy the condition $\delta \oint k_{\epsilon_i}=0$ in each case, where the variation $\delta$ becomes an exterior derivative on the reduced phase space.  Certainly, to have a true first law much more should be said, and it has been said, regarding the physical interpretation of each term, but the stress here is that the quantity sometimes playing the r\^ole of the mass can be relegated and be indirectly defined, in particular when the asymptotic time translation symmetry is not present or is difficult to identify.\footnote{For instance, this is the strategy used in \cite{Astorino:2016hls}, where the embedding of a charged and rotating black hole in a magnetic field makes subtle the selection of a preferred asymptotic time-like Killing vector field to define the spacetime mass.} 
 To decide the true thermodynamic value of the quasilocal first law relation obtained we would need to figure out a thermodynamics processes that allow us to change the value of the integrated charges. That is, a physical exchange of the amount of charges to flow in a description outside the reduced phase space, even when the usual far away of the black hole notion is not available. We leave this interesting question for future discussions.
 
Now, we give another step to generalize the formulae found for the surface charges.

\section {General Relativity and Electromagnetism in a D world}
\label{dworld}

In this section we further extend the applicability of surface charges by exhibiting the fundamental formulae for a $D$ dimensional manifold. We also show that for a particular case there is an explicit equivalence of the formulae of this approach with the recent ones worked in \cite{Barnich:2016rwk} with a different method.

For a spacetime of arbitrary dimension the Lagrangian of General Relativity coupled to Electromagnetism, with an a priori vanishing torsion, admits a generalization
\be
S[e,\omega,A]=\int_{{\cal M}}\left(\sum_{p=0}^{[D/2]}L^D_p +\alpha F*F\right),
\ee
where $L^D_p$ is a $D$-form given by
\be 
L_p^D=\alpha_p\varepsilon_{a_1\cdots a_D}R^{a_1a_2}\cdots R^{a_{2p-1}a_{2p}}e^{a_{2p+1}}\cdots e^{a_D},
\ee
the indexes are Lorentz and run as $a_1,\cdots,a_D=1,\cdots, D$.  The gravity part is known as the Lovelock action and for $D\neq 4$, the $\alpha_p$ are arbitrary coefficients.\footnote{For odd dimensions the Lovelock Lagrangian can be written as a Chern-Simons action for a particular fixation of the parameters \cite{Zanelli:2015pxa}.} If torsion is allowed more terms should be included in the action. We do not consider this further generalization with torsion because we do not know a compact way to treat all of them at once \cite{Mardones:1990qc}.

To get the surface charge we can either use the Noether approach detailed in section \ref{surfacecharges}, or the contracting homotopy operator (\ref{homotopy}).  The gravity contribution to the surface charge integrand obtained with the  Noether approach (\ref{SC}) is 
\ba
k^{GR}_\epsilon&=&-\sum_{p=1}^{\left[D/2\right]}\alpha_pp\,\varepsilon_{a_1\cdots a_D}\left(\lambda^{a_1a_2}\delta-\delta\omega^{a_1a_2}\xi\inter\right) \left(R^{a_3a_4}\cdots R^{a_{2p-1}a_{2p}} e^{a_{2p+1}}\cdots e^{a_D}\right),
\label{klovelock}
\ea
where $\delta$ and $\xi\inter$ act on the forms at the right. The formula coincides with the one obtained in \cite{Setare:2015cvv}. Operating with them we note that each $p$-term of the sum can be rewritten as
\begin{multline}
(D-2p)\varepsilon\left\{R\cdots R \left(\lambda\delta e-\delta\omega\xi\inter e\right)  e\cdots e\right\}\\ +(p-1)\varepsilon\left\{d\left(\lambda\delta\omega R\cdots Re\cdots e\right)+\delta_\epsilon \omega \delta \omega R\cdots Re\cdots e\right\},
\end{multline}
where the indexes are implicit to avoid cluttering. The terms in the second curly brackets are the generalization of (\ref{keuler}) to more dimensions. They are all exact forms plus a term proportional to the exactness condition. Therefore, they do not contribute to the surface charges. This property defines an equivalence relation among the surface charge integrands.
In particular, we can use the first terms of the previous equation to define
\ba
k'^{GR}_\epsilon&=&-\sum_{p=1}^{\left[\frac{D-1}{2}\right]}\alpha_pp(D-2p) \,\varepsilon_{a_1\cdots a_D} \left(\lambda^{a_{1}a_{2}}\delta e^{a_{3}}-\delta\omega^{a_{1}a_{2}}\xi\inter e^{a_{3}}\right)R^{a_4a_5}\cdots R^{a_{2p}a_{2p+1}} e^{a_{2p+2}}\cdots e^{a_D},\quad\quad
\label{klovelock2}
\ea
this would produce exactly the same surface charge than $k^{GR}_\epsilon$. Consequently, both belong to the same equivalence class, $k^{GR}_\epsilon\sim k'^{GR}_\epsilon$. Remarkably, the last expression is exactly the surface charge integrand computed directly with the contracting homotopy operator (\ref{homotopy})
\ba
k'^{GR}_\epsilon&=&I_{\delta e,\delta\omega}S_\epsilon.
\ea
Now, we choose to keep just the Einstein-Hilbert term in an arbitrary dimension, i.e., to keep only the $p=1$ term in the Lovelock action. Note that the cosmological constant term never contributes. The Einstein-Hilbert gravity contribution to the surface charge integrand (\ref{klovelock2}) in $D$ dimensions is
\be
 k^{EH}_\epsilon=-\alpha_1\left\{\lambda^{a_1a_2}\varepsilon_{a_1\cdots a_D}\delta \left(e^{a_{3}}\cdots e^{a_D}\right)-\delta\omega^{a_1a_2}\varepsilon_{a_1\cdots a_D}\xi\inter\left(e^{a_{3}}\cdots e^{a_D}\right)\right\},
\label{barnichetal}
\ee
notably it coincides  with the expression derived in \cite{Barnich:2016rwk}.\footnote{To do the comparison all quantities should be expanded in components and it should be noted the different prescription for the Lorentz gauge parameter $\delta_\epsilon=\Lie_\xi +\delta_{\bar\lambda}=\Lie_\xi +\delta_{\lambda+\xi\inter\omega}$, thus,  $\lambda^{a_1a_2}=\bar\lambda^{a_1a_2}-\xi\inter\omega^{a_1a_2}$. As explained before both prescriptions are equivalent, but the one we use produces formulae manifestly Lorentz invariant.}

This result tells us that the surface charges defined through the conventional symplectic method and the surface charges defined through the homotopy operator are equivalent. A formal proof of the last statement is done in the appendix \ref{appendix}. Furthermore, in \cite{Barnich:2016rwk} it is shown that the surface charges integrand formula, obtained for the tetrad-connection variables, is equivalent to the one written in pure metric variables (see for instance \cite{Barnich:2007bf}). Therefore, the formulae shown here (\ref{klovelock2}) are the natural generalization to an arbitrary dimension when all the Lovelock terms are considered either using tetrad-connection or metric variables.

The electromagnetic contribution computed with the Noether approach is direct. It is the trivial generalization to $D$ dimensions of the one obtained in (\ref{kGREM})
\be
k^{EM}_\epsilon = -2\alpha\left(\lambda \ \delta * F- \delta A\ \xi\inter* F\right),
\ee
it is naturally a $(D-2)-$form.

Therefore,  the total surface charge for gravity coupled to Electromagnetism in a $D$ world is
\be
k_\epsilon=k'^{GR}_\epsilon+k^{EM}_\epsilon.
\ee

The ensemble of these formulae allows us to define the surface charges associated for a large group of theories. For a given dimension $D$ one can pick any particular combination of Lovelock terms and couple it (or not) to electromagnetism. If such a theory have a well-defined family of solution with exact symmetries, then it is possible to define surface charges for them. The next step will be to integrate those surface charges to have finite charges. This can be accomplished by solving que condition $\delta \slashed \delta Q_\epsilon=0$ with the help of the remaining freedom in the parameters $\epsilon$. Those charges are the true physical quantities the phase space solution should be described with.

\section{Discussion}
\label{discussion}

Covariant symplectic methods is a powerful tool to deal with physical symmetries in gauge theories. The approach is old and spread, however several subtleties are usually disregarded (the triviality of a gauge current or the $\Theta([\delta_1,\delta_2])$ term, to point out some). One of the aims of these notes is to fill a key gap by obtaining the formulae of the surface charges from the usual canonical symplectic approach (\ref{holahola}). Another aim is to show the relation with the so called {\it invariant} symplectic approach based on the contracting homotopy operators \cite{Barnich:2007bf}. We do so because a part of the community is unaware of the powerful results related with the surface charges. And specifically, unaware of the way they help to solve the problems of using Noether currents in gauge theories.

It is a result that both symplectic approaches to define surface charges are equivalent as far as the assumptions to build the charges are respected. This is shown in the appendix \ref{appendix} for the general case and it was checked in section \ref{dworld} for the theory of gravity in an arbitrary dimension.
In this regard the moral from both approaches is that for gauge theories physical symmetries are better understood at the level of the symplectic structure, not at the level of Noether currents. It is at the symplectic structure level where the conservation of surface charges can be established.

On the same line we foment the use of surface charges in gravity by deriving their explicit formulae for the first order formulation of General Relativity coupled to Electromagnetism, based on tetrad and connection variables, (\ref{kGREM}). This is one of our main results. We further extend this result for an arbitrary dimension, (\ref{klovelock}). The elegance of the language is expressed in the simplicity of the formulae obtained. No reference to coordinates is required. On the other hand, the translation to components depending on coordinates is straightforward as it was stated in the comparison of (\ref{barnichetal}) with \cite{Barnich:2016rwk}.

In subsection \ref{bhexample}  we applied the formalism to the 3+1 solution family of electrically charged, rotating, and asymptotically constant curvature black holes. We also exploited the quasilocal nature of the surface charges to present an alternative way to recover the first law of black hole mechanics. 

Another interesting application of the surface charges formalism can be done in 2+1 dimensions for the exact symmetries of the electrically charged BTZ black hole. The computation can be contrasted with the general asymptotic approach developed in \cite{Perez:2015jxn}. In \cite{Perez:2015jxn} the standard Brown-Henneaux boundary condition were relaxed such that the Hamiltonian analysis could include logarithmic divergent tailing of the fields, as the ones presented in the charged BTZ black hole. The resulting formulae for the {\it surface integrals} (Regge-Teitelboim method) is applied to the charged BTZ black hole solution. The first law is satisfied by those charges. Based on partial computation we advance that using the surface charges approach computed in section \ref{dworld} for $D=3$ the result is the same.

Another interesting application of the machinery just presented is on the so-called {\it asymptotic symmetries}. Let us briefly describe the program to set the questions properly. The phase space of General Relativity can be explored by perturbing the fields around an arbitrary solution. A whole research program is built on ways to perform field perturbations around solutions such that they describe a large class of spacetimes. For instance, field perturbations are usually encoded in the so-called boundary conditions: Particular tailing for the fields far away from sources (e.g. the definition of asymptotically flat, de Sitter, or anti-de Sitter spacetimes). There is an interesting game in the way those boundary conditions are specified such that they reflect one or other physical situation (for example in asymptotically flat spacetimes to allow or not radiation at future null infinity). Given any particular solution, the same strategy can be applied in a specific spacetime region. Then, by studying the symmetries of this enlarged family of spacetimes it is possible to define a larger group of symmetries than the one of the starting unperturbed solution. The surface charge formulae are quasilocal and thus they can be used for those asymptotic symmetries producing non-vanishing quantities. In fact non-vanishing surface charges could be computed even for gauge transformation which are not physical symmetries in the sense that do not respect the exactness condition. Then, the main question is to decide if the so-called asymptotic symmetries have a physical meaning. In other words, if different values of the surface charges associated with asymptotic symmetries correspond to physically different solutions. It could happen that the enlarged group of symmetries have a lot of gauge redundancy and in consequence their surface charges belong to the same equivalence class. Another possibility is that they are truly physical. In this case, the phase space of spacetime solutions would have a long time disregarded and richer structure.

Assuming the later possibility we have for instance \cite{Hawking:2016msc}. It is pointed out that an infinite number of {\it charges} associated to asymptotic symmetries, defined on the null region of an asymptotically flat spacetime, are related to an infinite number of {\it charges} defined for near horizon symmetries. For the charges defined in the family of black hole solutions studied in this note, the (anti-)de Sitter Kerr-Newman family, the statement is trivial because of the surface charges conservation (note that the asymptotically flat family is treated exactly in the same way). In this regard, and assuming that the asymptotic symmetries are exact, it would be interesting to find a systematic way to perturb this solution family, and therefore to define a larger family of solutions, such that a larger group of exact symmetries can be defined for it in the whole spacetime. Then, the associated surface charges could be computed at once either on closed two-surfaces at the asymptotic region, or at the near horizon region (actually on any two-surface enclosing the singularity). In this program part of the machinery is already at disposal but what is missing is the detailed description of such a family of perturbed solutions and its exact symmetries.

To study the surface charges associated to the asymptotic symmetries in the tetradic and connection variables is one of the future directions of this work. The outcome should be a better understanding of the physical relevance of those constructions.

If the outcome is positive, that is, if there exist such a description of perturbed solutions admitting an infinite number of exact symmetries and related charges, the expectation is that the extra symmetries are generic for all spacetimes, regardless of their particular asymptotic structure, and in fact a quasilocal property.

To decide the value of the previous ideas the covariant symplectic formalism and the surface charges expressed with tetradic and connection variables offers a solid starting point. 

\newpage

\section*{Acknowledgments}

The authors thank Alfredo Guevara for collaboration at the early stages of this project.

The authors are grateful to G. Barnich, G. Comp\`ere, A. Kegel, C. Troessaert,  and J. Zanelli for enlightening discussions. EF  wishes to thank the Quantum Gravity division of the Max Planck Institute (Germany) and the Instituto de F\'isica of Universidad de la Rep\'ublica (Uruguay) for the hospitality and support during his research visits. EF is partially funded by Fondecyt grant 11150467. DH is partially founded by Conicyt grant 21160649.  The Centro de Estudios Cient\'ificos (CECs) is funded by the Chilean Government through the Centers of Excellence Base Financing Program of Conicyt.

\appendix

\section{Comparison of surface charge definitions}
\label{appendix}

To make contact with other approaches, in this section we introduce a different definition for surface charges used in \cite{Barnich:2007bf}, and further import the comparison with the prescription presented in section \ref{surfacecharges}. The key of this different definition is its direct use of $S_{\epsilon}$ introduced in (\ref{noether0}), that is, the particular equation of motions combined with the gauge parameters that results from the use of Noether identities. In other words, the only term appearing in the trivially conserved current, $J_\epsilon=\Theta(\delta_\epsilon\Phi)-\xi\inter L+S_\epsilon$, that does not depend directly of the Lagrangian boundary term.
The surface charge integrand is expressed as
\ba
k'_\epsilon\equiv I_{\delta\Phi}S_\epsilon,
\ea
where $I_{\delta\Phi}$ is called the homotopy operator. The homotopy operator is an efficient way to get a sensible $(p-1)$-form from an exact $p$-form. In particular it can be used to select the boundary term in the Lagrangian variation
\be
\delta L=E\delta\Phi+d\left[\Theta'(\delta\Phi)+dY\right]=E\delta\Phi+d\left[I_{\delta\Phi}L\right].
\label{boundaryfixed}
\ee
With the risk of keeping the discussion rather abstract while brief, we just pick up the properties that allow us to understand the comparison (see \cite{Barnich:2007bf} for a detailed definition of the homotopy operator).  The definining property of the homotopy operator is its relation with a variation of fields in the space of configuration
\ba
\delta'\equiv d I_{\delta\Phi}+I_{\delta\Phi}d,
\label{deltaprima}
\ea
where $d$ is the exterior derivative. In fact the homotopy operator provides a prescription to define a variation  on the phase space. Therefore we called it $\delta'$ to distinguish it from our treatment. Note the analogy with the expression of the spacetime Lie derivative (\ref{cartanformula}). 

Already with this property we can prove
\ba
dk'_\epsilon&=&-I_{\delta\Phi}dS_\epsilon+\delta' S_\epsilon\\
&=&-I_{\delta\Phi}[E\delta_\epsilon \Phi]+\delta' S_\epsilon\\
&=&-I_{\delta\Phi}[E]\delta_\epsilon \Phi-(-1)^{p_E} E\ I_{\delta\Phi}[\delta_\epsilon \Phi]+\delta' S_\epsilon,
\ea
where we used the Noether identities $E\delta_\epsilon \Phi=dS_\epsilon-N_\epsilon=dS_\epsilon$, and $p_E$ is the form degree of $E$, i.e., $I_{\delta\Phi}E=(-1)^{p_E}E\ I_{\delta\Phi}$. Therefore, it is shown that $k'_{\epsilon}$ is  closed if  the equation of motion, the linearized equation of motions, and the exactness condition hold, i.e., $E=0$, $\delta E=0$, and $\delta_\epsilon\Phi=0$. These conditions are exactly the ones required for the surface charge integrand defined in (\ref{kconserved}) to be closed. In the previous calculation we made use of the so-called {\it invariant} presymplectic structure density
\ba
\Omega'(\delta_1,\delta_2)\equiv I_{\delta_{[1}\Phi}(E\delta_{2]}\Phi).
\ea
It differs from the presymplectic structure density introduced before 
\ba
\Omega(\delta_1,\delta_2)=\delta_1\Theta(\delta_2\Phi)-\delta_2\Theta(\delta_1\Phi)-\Theta([\delta_1,\delta_2]\Phi).
\ea
Both prescription are in general inequivalent as it is shown in the following.

The boundary term $\Theta(\delta\Phi)$ has an intrinsic ambiguity that can be selected with the homotopy operator (\ref{boundaryfixed}), we use it to fix the ambiguity of the presymplectic structure density
\be
\Omega(\delta_1,\delta_2)=\delta'_1( I_{\delta_{2}\Phi}L)-\delta'_2( I_{\delta_{1}\Phi}L).
\ee
The use of $\delta'_{1,2}$ as defined by (\ref{deltaprima}) ensure linearity in the variations, then to introduce the commutator term is unnecessary.
Although we have selected the boundary term, there is still another intrinsic ambiguity if the Langrangian is allowed to change by an exact form, $L\to L+d\alpha$, it is in this sense that this prescription for the symplectic structure density is not {\it invariant}. The comparison of both presymplectic structure densities goes as
\ba
\Omega'(\delta_1,\delta_2)&=&I_{\delta_{[1}\Phi}(E\delta_{2]} \Phi)\\
&=&I_{\delta_{[1}\Phi}(\delta'_{2]} L-dI_{\delta_{2]}\Phi}L)\\
&=&\delta'_{[2}I_{\delta_{1]}\Phi}L+\delta'_{[1}I_{\delta_{2]}\Phi}L-d\left(I_{\delta_{[1}\Phi}I_{\delta_{2]}\Phi}L\right)\\
&=&\Omega(\delta_1,\delta_2)-d\widetilde E_{1,2}
\ea
where we used that  the homotopy operator satisfies $I_{\delta_{1}}\delta'_2=\delta'_2I_{\delta_1}$,\footnote{In this notation this property is the equivalent of equation $[d_v,I_{d_v}]=0$ where $d_v$ denotes {\it vertical derivatives} (see A.5 in \cite{Barnich:2007bf}).} and $\widetilde E_{1,2}\equiv I_{\delta_{[1}\Phi}I_{\delta_{2]}\Phi}L$. Thus, in the case we have exact symmetries, $\widetilde E$ vanishes and there is a match in both prescriptions. This is the generalization of what it was described in the computation of section \ref{dworld}. 

It is worth to point out the differences in the prescription: $k'_\epsilon$ and $\Omega'(\delta_1,\delta_2)$ depend directly on the equation of motions and it is insensitive to the intrinsic ambiguities of the variational principle. On the other hand, $k_\epsilon$ and $\Omega(\delta_1,\delta_2)$ can be computed from standard procedures without introducing the homotopy operator.  As a final remark, we note that in (\ref{homotopy}) we exhibited and explicit formula for the homotopy operator written for a gravity theory in tetrad-connection variables.


\begin{thebibliography}{100}

\bibitem{Ashtekar:2000eq}
  A.~Ashtekar, J.~C.~Baez and K.~Krasnov,
  ``Quantum geometry of isolated horizons and black hole entropy,''
  Adv.\ Theor.\ Math.\ Phys.\  {\bf 4} (2000) 1
  \href{https://arxiv.org/pdf/gr-qc/0005126.pdf}{[gr-qc/0005126]}
  
\bibitem{Engle:2009vc}
  J.~Engle, A.~P\'erez and K.~Noui,
  ``Black hole entropy and SU(2) Chern-Simons theory,''
  Phys.\ Rev.\ Lett.\  {\bf 105} (2010) 031302
  \href{https://arxiv.org/pdf/0905.3168.pdf}{[gr-qc/0905.3168]}
    
\bibitem{Engle:2010kt}
  J.~Engle, K.~Noui, A.~P\'erez and D.~Pranzetti,
  ``Black hole entropy from an SU(2)-invariant formulation of Type I isolated horizons,''
  Phys.\ Rev.\ D {\bf 82} (2010) 044050
   \href{https://arxiv.org/pdf/1006.0634.pdf}{[gr-qc/1006.0634]}
  
 \bibitem{Perez:2010pq}
  A.~P\'erez and D.~Pranzetti,
  ``Static isolated horizons: SU(2) invariant phase space, quantization, and black hole entropy,''
  Entropy {\bf 13} (2011) 744
  \href{https://arxiv.org/pdf/1011.2961.pdf}{[gr-qc/1011.2961]}
    
\bibitem{Brown:1986nw}
  J.~D.~Brown and M.~Henneaux,
  ``Central Charges in the Canonical Realization of Asymptotic Symmetries: An Example from Three-Dimensional Gravity,''
  Commun.\ Math.\ Phys.\  {\bf 104} (1986) 207
  \href{https://doi.org/10.1007/BF01211590}{doi:10.1007/BF01211590}

\bibitem{Barnich:2001jy}
  G.~Barnich and F.~Brandt,
  ``Covariant theory of asymptotic symmetries, conservation laws and central charges,''
  Nucl.\ Phys.\ B {\bf 633} (2002) 3
  \href{https://arxiv.org/pdf/hep-th/0111246.pdf}{[hep-th/0111246]}

\bibitem{Strominger:1997eq}
  A.~Strominger,
  ``Black hole entropy from near horizon microstates,''
  JHEP {\bf 9802} (1998) 009
  \href{https://arxiv.org/pdf/hep-th/9712251.pdf}{[hep-th/9712251]}

 \bibitem{Guica:2008mu}
  M.~Guica, T.~Hartman, W.~Song and A.~Strominger,
  ``The Kerr/CFT Correspondence,''
  Phys.\ Rev.\ D {\bf 80} (2009) 124008
  \href{https://arxiv.org/pdf/0809.4266.pdf}{[hep-th/0809.4266]}
  
\bibitem{Compere:2015mza}
  G.~Comp\`ere, K.~Hajian, A.~Seraj and M.~M.~Sheikh-Jabbari,
  ``Extremal Rotating Black Holes in the Near-Horizon Limit: Phase Space and Symmetry Algebra,''
  Phys.\ Lett.\ B {\bf 749} (2015) 443
  \href{https://arxiv.org/pdf/1503.07861.pdf}{[hep-th/1503.07861]}
   
\bibitem{Afshar:2016uax}
  H.~Afshar, D.~Grumiller and M.~M.~Sheikh-Jabbari,
 ``Black Hole Horizon Fluffs: Near Horizon Soft Hairs as Microstates of Three Dimensional Black Holes,''
   \href{https://arxiv.org/pdf/1607.00009.pdf}{[hep-th/1607.00009]}

  
\bibitem{Hawking:2016msc}
  S.~W.~Hawking, M.~J.~Perry and A.~Strominger,
  ``Soft Hair on Black Holes,''
  Phys.\ Rev.\ Lett.\  {\bf 116} (2016) no.23,  231301
   \href{https://arxiv.org/pdf/1601.00921.pdf}{[hep-th/1601.00921]}

\bibitem{Carlip:1995zj}
  S.~Carlip,
  ``Lectures on (2+1) dimensional gravity,''
  J.\ Korean Phys.\ Soc.\  {\bf 28} (1995) S447
  \href{https://arxiv.org/pdf/gr-qc/9503024.pdf}{[gr-qc/9503024]}


\bibitem{Geiller:2017xad}
  M.~Geiller,
  ``Edge modes and corner ambiguities in 3d Chern-Simons theory and gravity,''
  \href{https://arxiv.org/pdf/1703.04748.pdf}{[gr-qc/1703.04748]}

\bibitem{Compere:2007az}
  G.~Comp\`ere,
  ``Symmetries and conservation laws in Lagrangian gauge theories with applications to the mechanics of black holes and to gravity in three dimensions,''
  \href{https://arxiv.org/pdf/0708.3153.pdf}{[hep-th/0708.3153]}

\bibitem{Barnich:2016rwk}
  G.~Barnich, P.~Mao and R.~Ruzziconi,
  ``Conserved currents in the Cartan formulation of general relativity,''
\href{https://arxiv.org/pdf/1611.01777.pdf}{[gr-qc/1611.01777]}

\bibitem{Wald:1999wa}
  R.~M.~Wald and A.~Zoupas,
  ``A General definition of 'conserved quantities' in general relativity and other theories of gravity,''
  Phys.\ Rev.\ D {\bf 61} (2000) 084027
  \href{https://arxiv.org/pdf/gr-qc/9911095.pdf}{[gr-qc/9911095]}
  
\bibitem{Iyer:1994ys}
  V.~Iyer and R.~M.~Wald,
  ``Some properties of N\"other charge and a proposal for dynamical black hole entropy,''
  Phys.\ Rev.\ D {\bf 50} (1994) 846
  \href{https://arxiv.org/pdf/gr-qc/9403028.pdf}{[gr-qc/9403028]}
  
\bibitem{Barnich:2007bf}
  G.~Barnich and G.~Comp\`ere,
  ``Surface charge algebra in gauge theories and thermodynamic integrability,''
  J.\ Math.\ Phys.\  {\bf 49} (2008) 042901
  \href{https://arxiv.org/pdf/0708.2378.pdf}{[gr-qc/0708.2378]}
  
\bibitem{Compere:2009dp}
  G.~Comp\`ere, K.~Murata and T.~Nishioka,
  ``Central Charges in Extreme Black Hole/CFT Correspondence,''
  JHEP {\bf 0905} (2009) 077
  \href{https://arxiv.org/pdf/0902.1001.pdf}{[hep-th/0902.1001]}


\bibitem{Aros:1999id}
  R.~Aros, M.~Contreras, R.~Olea, R.~Troncoso and J.~Zanelli,
  ``Conserved charges for gravity with locally AdS asymptotics,''
  Phys.\ Rev.\ Lett.\  {\bf 84} (2000) 1647
  \href{https://arxiv.org/pdf/gr-qc/9909015.pdf}{[gr-qc/9909015]}
   
\bibitem{Green:2013ica}
  S.~R.~Green, J.~S.~Schiffrin and R.~M.~Wald,
  ``Dynamic and Thermodynamic Stability of Relativistic, Perfect Fluid Stars,''
  Class.\ Quant.\ Grav.\  {\bf 31} (2014) 035023
   \href{https://arxiv.org/pdf/1309.0177.pdf}{[gr-qc/1309.0177]}
 
\bibitem{Tachikawa:2006sz}
  Y.~Tachikawa,
  ``Black hole entropy in the presence of Chern-Simons terms,''
  Class.\ Quant.\ Grav.\  {\bf 24} (2007) 737
 \href{https://arxiv.org/pdf/hep-th/0611141.pdf}{[hep-th/0611141]}
\bibitem{Prabhu:2017}
K. Prabhu,
``The First Law of Black Hole Mechanics for Fields with Internal Gauge Freedom,''
Class.\ Quant.\ Grav.\  {\bf 34} (2017) no3, 035011
\href{https://arxiv.org/pdf/1511.00388.pdf}{[hep-th/1511.00388]}

\bibitem{Barnich:2000zw}
  G.~Barnich, F.~Brandt and M.~Henneaux,
  ``Local BRST cohomology in gauge theories,''
  Phys.\ Rept.\  {\bf 338} (2000) 439
  \href{https://arxiv.org/pdf/hep-th/0002245.pdf}{[hep-th/0002245]}

\bibitem{Zanelli:provisory}
  M.~Hassaine and J.~Zanelli, ``Chern-Simons (Super)Gravity,''  100 Years of General Relativity - Vol. 2,  World Scientific Publishing (2016)

\bibitem{MM} S.\ W.\ MacDowell and F.\ Mansouri, ``Unified geometric theory of gravity and supergravity,'' {\sl Phys.\ Rev.\ Lett.\ }{\bf 38}
(1977), 739--742.  [Erratum, {\sl ibid.} {\bf 38} (1977), 1376]
\href{https://doi.org/10.1103/PhysRevLett.38.739}{doi:10.1103/PhysRevLett.38.739}

\bibitem{Wise:2006sm}
  D.~K.~Wise,
  ``MacDowell-Mansouri gravity and Cartan geometry,''
  Class.\ Quant.\ Grav.\  {\bf 27} (2010) 155010
    [gr-qc/0611154]
   \href{https://arxiv.org/pdf/gr-qc/0611154.pdf}{[gr-qc/0611154]}
  
\bibitem{Ashtekar:2008jw}
  A.~Ashtekar, J.~Engle and D.~Sloan,
  ``Asymptotics and Hamiltonians in a First order formalism,''
  Class.\ Quant.\ Grav.\  {\bf 25} (2008) 095020
 \href{https://arxiv.org/pdf/gr-qc/0802.2527.pdf}{[gr-qc/0802.2527]}

\bibitem{Jacobson:2015uqa}
  T.~Jacobson and A.~Mohd,
  ``Black hole entropy and Lorentz-diffeomorphism N\"other charge,''
  Phys.\ Rev.\ D {\bf 92} (2015) 124010
   \href{https://arxiv.org/pdf/1507.01054.pdf}{[gr-qc/1507.01054]}

\bibitem{Montesinos:2017epa}
  M.~Montesinos, D.~González, M.~Celada and B.~Díaz,
  ``Reformulation of the symmetries of first-order general relativity,''
  Class.\ Quant.\ Grav.\  {\bf 34} (2017) no.20,  205002
   \href{https://arxiv.org/pdf/1704.04248.pdf}{[gr-qc/1704.04248]}

\bibitem{Henneaux:1992ig}
  M.~Henneaux and C.~Teitelboim,
  ``Quantization of gauge systems,''
  Princeton, USA: Univ. Pr. (1992) 520 p

\bibitem{Corichi:2016zac}
  A.~Corichi, I.~Rubalcava-García and T.~Vukašinac,
  ``Actions, topological terms and boundaries in first-order gravity: A review,''
  Int.\ J.\ Mod.\ Phys.\ D {\bf 25} (2016) no.04,  1630011
  \href{https://arxiv.org/pdf/1604.07764.pdf}{[gr-qc/1604.07764]}

\bibitem{Banados:2016zim}
  M.~Bañados and I.~A.~Reyes,
  ``A short review on Noether’s theorems, gauge symmetries and boundary terms,''
  Int.\ J.\ Mod.\ Phys.\ D {\bf 25} (2016) no.10,  1630021
  \href{https://arxiv.org/pdf/1601.03616.pdf}{[hep-th/1601.03616]}

\bibitem{Barnich:2003xg}
  G.~Barnich,
  ``Boundary charges in gauge theories: Using Stokes theorem in the bulk,''
  Class.\ Quant.\ Grav.\  {\bf 20} (2003) 3685
  \href{https://arxiv.org/pdf/hep-th/0301039.pdf}{[hep-th/0301039]}

\bibitem{Astorino:2016hls}
  M.~Astorino, G.~Comp\`ere, R.~Oliveri and N.~Vandevoorde,
  ``Mass of Kerr-Newman black holes in an external magnetic field,''
  Phys.\ Rev.\ D {\bf 94} (2016) no.2,  024019
  \href{https://arxiv.org/pdf/1602.08110.pdf}{[gr-qc/1602.08110]}

\bibitem{Caldarelli:1999xj}
  M.~M.~Caldarelli, G.~Cognola and D.~Klemm,
  ``Thermodynamics of Kerr-Newman-AdS black holes and conformal field theories,''
  Class.\ Quant.\ Grav.\  {\bf 17} (2000) 399
   \href{https://arxiv.org/pdf/hep-th/9908022.pdf}{[hep-th/9908022]}
 
\bibitem{Zanelli:2015pxa}
  J.~Zanelli,
  ``Chern-Simons Forms and Gravitation Theory,''
  Lect.\ Notes Phys.\  {\bf 892} (2015) 289,
  \href{https://doi.org/10.1007/978-3-319-10070-8_11}{doi:10.1007/978-3-319-10070-8 11}

    
\bibitem{Mardones:1990qc}
  A.~Mardones and J.~Zanelli,
  ``Lovelock-Cartan theory of gravity,''
  Class.\ Quant.\ Grav.\  {\bf 8} (1991) 1545,
    \href{https://doi.org/10.1088/0264-9381/8/8/018}{doi:10.1088/0264-9381/8/8/018}

\bibitem{Setare:2015cvv}
  H.~Adami and M.~R.~Setare,
  ``Quasi-local conserved charges in Lorenz–diffeomorphism covariant theory of gravity,''
  Eur.\ Phys.\ J.\ C {\bf 76} (2016) no.4,  187
\href{https://arxiv.org/pdf/1511.00527.pdf}{[gr-qc/1511.00527]}


\bibitem{Perez:2015jxn}
  A.~Perez, M.~Riquelme, D.~Tempo and R.~Troncoso,
  ``Asymptotic structure of the Einstein-Maxwell theory on AdS$_{3}$,''
  JHEP {\bf 1602} (2016) 015
\href{https://arxiv.org/pdf/1512.01576.pdf}{[hep-th/1512.01576]}


\end{thebibliography}

\end{document}